\documentstyle[12pt]{article}

\input psfig.sty

\newcommand{\be}{\begin{equation}}
\newcommand{\ee}{\end{equation}}
\newcommand{\bea}{\begin{eqnarray}}
\newcommand{\eea}{\end{eqnarray}}
\newcommand{\rp}{{\rm peak}}
\newcommand{\fm}{{\rm fm}}
\newcommand{\rpp}{{\rm p}}

\def\NPB{{Nucl. Phys.} B}
\def\PLB{{Phys. Lett.} B}
\def\PRL{Phys. Rev. Lett.}
\def\CMP{Comm. Math. Phys.}

\def\PRD{{Phys. Rev.} D}

\newcommand{\basispl}{
   \put(-1.,-.5){\line(1,0){2}}
   \put(-1.,.5){\line(1,0){2}}
   \put(-1.,-.5){\line(0,1){1}}
   \put(1.,-.5){\line(0,1){1}}
                         }
\newcommand{\basispr}{
   \put(-.5,-1.){\line(1,0){1}}
   \put(-.5,1.){\line(1,0){1}}
   \put(-.5,-1.){\line(0,1){2}}
   \put(.5,-1.){\line(0,1){2}}
                         }

\newcommand{\basisal}{
   \put(.0,.5){\vector(1,0){0}}
   \put(.0,-.5){\vector(-1,0){0}}
   \put(-1,.0){\vector(0,1){0}}
   \put(1,.0){\vector(0,-1){0}}
                         }   
\newcommand{\basisar}{
   \put(.0,1.){\vector(1,0){0}}
   \put(.0,-1.){\vector(-1,0){0}}
   \put(-.5,.0){\vector(0,1){0}}
   \put(.5,.0){\vector(0,-1){0}}
                      }

\newcommand{\cloverl}{\setlength{\unitlength}{.5cm}\raisebox{-.5cm}
{   \begin{picture}(2.8,2.8)(-1.6,-1.2)
   \multiput(-1.2,-0)(2.2,0.){2}{\begin{picture}(1.2,1.2)(-.6,-.6)
   \basispl \basisal
   \end{picture}}
   \multiput(-1.2,-1.2)(2.2,0.){2}{\begin{picture}(1.2,1.2)(-.6,-.6)
   \basispl \basisal
   \end{picture}}
   \put(.4,-.1){\circle*{.2}}
   \put(.4,.1){\circle*{.2}}
   \put(.65,-.1){\circle*{.2}}
   \put(.65,.1){\circle*{.2}}
   \end{picture}}}
\newcommand{\cloverr}{\setlength{\unitlength}{.5cm}\raisebox{-.5cm}
{   \begin{picture}(2.8,2.8)(-1.6,-1.6)
   \multiput(-2.2,0)(1.2,0){2}{\begin{picture}(1.2,1.2)(-.6,-.6)
   \basispr \basisar  
   \end{picture}}  
   \multiput(-2.2,-2.2)(1.2,0){2}{\begin{picture}(1.2,1.2)(-.6,-.6)
   \basispr \basisar  
   \end{picture}}  
   \put(-.9,-.6){\circle*{.2}}
   \put(-.9,-.4){\circle*{.2}}
   \put(-1.1,-.6){\circle*{.2}}
   \put(-1.1,-.4){\circle*{.2}}
   \end{picture}}}
\newcommand{\twooneplaq}{\setlength{\unitlength}{.5cm}
   \raisebox{-.2cm}{
   \begin{picture}(2.2,1.2)(-1.0,-.6)
   \basispl\basisal
   \put(-1.6,-0.9){$x$}
   \put(1.2,-0.9){$x\!+\!m\mu$}
   \put(-2.3,0.9){$x\!+\!n\nu$}
   \put(1,-.5){\circle*{.2}}
   \put(-1,-.5){\circle*{.2}}
   \put(-1.,.5){\circle*{.2}}
   \end{picture}}}

\def\sqr#1#2#3{{\vcenter{\hrule height.#2pt
      \hbox{\vrule width.#2pt height#1pt \kern#3pt
         \vrule width.#2pt}
      \hrule height.#2pt}}}

\def\mystrut{{\vrule height 13pt depth 6pt width 0pt}}

\begin{document}

\vskip4mm
\begin{center}
{\LARGE{\bf{Topology of the SU(2) vacuum: a lattice study using 
improved cooling}}}\\
\vspace{1cm}
{\large Philippe  de  Forcrand${}^a$, Margarita Garc\'{\i}a P\'erez${}^b$ \\
and \\
Ion-Olimpiu  Stamatescu${}^{c,b}$
} \\
\vspace{1cm}
${}^a$ SCSC, ETH-Z\"urich, CH-8092 Z\"urich, Switzerland.\\
${}^b$ Institut f. Theoretische Physik, University of Heidelberg,\\
D-69120 Heidelberg, Germany.\\
${}^c$ FEST, Schmeilweg 5, D-69118 Heidelberg, Germany.\\
\end{center}

\vspace*{5mm}{\narrower\narrower{\noindent
\underline{Abstract:} We study the topological structure of
the SU(2) vacuum at zero temperature: 
topological susceptibility, size, shape and distance distributions of the
instantons.
We use a cooling  algorithm based on 
an improved action with scale invariant instanton solutions. 
This algorithm needs no monitoring
or calibration, has an inherent cut off for dislocations and leaves unchanged 
instantons at  physical scales. 
The physical relevance of our
results is checked by studying the scaling and finite volume dependence. 
We obtain a susceptibility of $(200(15){\rm MeV})^4$. 
The instanton size distribution is peaked around $0.43$fm, and the
distance distribution indicates a homogeneous, random spatial structure.
}
\par}
  
\section{ Introduction }

Much work has been devoted to the study of topology within the framework of lattice
gauge theories. Among other topics the measurement of the topological
susceptibility and investigations of the relevance of instantons
to the QCD vacuum have attracted a lot of interest. 
  
The topological susceptibility is a physical quantity which measures the fluctuations
of the topological charge of the vacuum. It is related via the Witten-Veneziano formula \cite{WI,VE}
to a phenomenological parameter (the $\eta' - \eta$ mass difference)
and attributed to the presence of instantons in the QCD vacuum.
It can be measured on the lattice where it is expected to scale
according to its dimension as the continuum limit is approached.  

Instantons are 
classical solutions of the (euclidean) equations of motion, that is,
they represent minima of the euclidean action \cite{bela,tho}. 
They have been conjectured to be relevant for various effects like the hadronic
spectrum or the chiral phase transition and used as the basic degrees of
freedom of several effective models.
Such models assume certain characteristics of the instanton ensemble 
concerning for instance the density correlations (crystal, liquid or gas), 
the size distribution and other features \cite{models}.
Tests of the assumptions on which these models are based, or direct tests of the 
physical effects of instantons can be provided by the study of the physical, 
non-trivial topological excitations  present in lattice configurations.

 In connection with the measurement of the topological susceptibility on
the lattice most approaches have concentrated on devising appropriate
topological charge operators. The operators commonly used belong to
three categories:
\begin{itemize}
\item
Geometric, based on a geometric construction of the topological
charge \cite{lues,ps}.
\item
Analytic, based on the field-theoretical definition of the charge
\cite{peskin, dV}. 
\item
Fermionic, based on the relation between the topological charge
and the zero modes of the Dirac operator \cite{tho,nucu,rossi,smit}.
\end{itemize}
The latter, although promising, has not been pursued too far yet 
due to the problems involved in taking the zero mass limit of the
Dirac operator. There has also been a recent proposal for simulating chiral 
fermions on the lattice (the overlap formalism) \cite{nara2} which simultaneously 
provides a method to measure the topological charge. It has been
applied to measure the topological susceptibility and preliminary results indicate  
good agreement with the Witten-Veneziano formula \cite{nara}. 

The other two approaches face several difficulties related to  lattice
artifacts due in particular to the bad
behavior of the Wilson action regarding scale-invariance. 
Instead of assigning to an instanton an action independent of its size,
the Wilson action gives an action which monotonically decreases with the
instanton size.  This
leads, in a Monte Carlo simulation, to an unphysical abundance of small
instantons at the scale of the cut-off (``dislocations").
These dislocations are not physical but still they contribute to the geometrical
charge which hence
tends to overestimate the topological susceptibility \cite{geom1,geom2}.
According to an argument by Pugh and Teper \cite{PT} their contribution will even   
diverge in the continuum limit.
On the other hand naive field-theoretical definitions of the topological charge 
have strong renormalization effects induced by the lattice
regularization which tend to decrease the topological susceptibility 
and again spoil its scaling behavior \cite{dg1}.

These problems can be handled at two levels: 
in the Monte Carlo simulation, by improving the scale invariance of 
the action used in order to decrease
the contribution from dislocations; or in the measurement,
by designing a method to disentangle dislocations from physical
instantons and to decrease renormalization effects.

Both approaches have been pursued. Cooling (an iterative minimization of the lattice action) \cite{cool} 
has been used to handle the problem at the level of measurement.
Ideally it  
eliminates rough topological fluctuations (dislocations) while keeping
large instantons  unchanged and decreases renormalization effects
by smoothing out the ultraviolet noise.
For the configurations remaining after cooling 
 good agreement is found between field-theoretical and geometric 
determinations of the topological charge.  However cooling with the Wilson
action induces more changes than just smoothing out the UV noise
and eliminating dislocations. The
decrease of the Wilson action with the instanton size means that this action
has actually no stable instantons: under cooling with the Wilson action
even wide (physically relevant) instantons shrink and decay.
Hence cooling algorithms based on such action cannot provide 
a reliable plateau  
for the topological charge as a function of cooling time.
Moreover these algorithms do not preserve the 
physical scale structure of the original configurations.
Because of this one usually attempts to calibrate or engineer
the cooling procedure based on the Wilson action so as to obtain some
meta-stability window \cite{dg2,MS},
after the noise has been reduced but before 
instantons have shrunk to zero. This, however, makes
cooling more an art than a method and also does not  unambiguously preserve
information concerning, for instance, the physical size of the
instantons present in the original configuration.

There have been quite successful attempts to combine cooling with other methods,
like ``heating" \cite{dg1}, to compute the lattice
renormalization factor of the topological susceptibility, or
more recently to define ``smeared" topological charge operators \cite{dg3}.
They were shown to improve  susceptibility data but no results have been provided 
concerning size distributions or other features of the topological structure of the vacuum.

At the level of the Monte Carlo simulations, improved actions, which generally
have
better scaling properties than Wilson action, are expected to at least reduce the
contribution of dislocations to the susceptibility \cite{PT}.
  Two proposals in this direction have been formulated recently.
On one hand ``fixed point perfect actions" have been constructed in terms of many loops and
higher representations \cite{Ha}. These actions are defined with the help of a blocking procedure whose choice
should ensure the persistence of the good scaling qualities found at the
fixed point even far from it, where the simulations are performed.
On the other hand a proposal for a lattice action has been formulated
using interpolating gauge fields,
with the aim of ensuring on the lattice some of the properties of
the continuum fields \cite{pilar}.
Both of these approaches could be argued at the semi-classical level
 not to give a divergent topological susceptibility in the continuum limit.
The first analyses of SU(2) topology using perfect actions
have already been performed \cite{HG}, \cite{Ilg}. They indicate that
even in this case there is an abundance of dislocations in the
Monte Carlo configurations.  To disentangle them from physically relevant
structures a refined construction in many steps, using  special reverse blocking
procedures in combination with minimization steps has been developed to take advantage of the qualities of the
fixed point actions \cite{HG}, \cite{Ilg}. By fixing the number of blockings
(typically, one) one fixes the scale (in lattice units) at which one wants to extract the topological structure.

We consider here the problem of measurement, although the lattice
operators we use could also be used for the Monte Carlo 
generation of configurations, with great benefits.
We propose to take advantage of the better scale-invariance
of improved actions to design
a cooling algorithm which by itself avoids most of the diseases of the usual cooling, and use it to analyze data from Monte Carlo simulations with any action. 
In fact, cooling should fulfill a number of requirements
in order to qualify as an IR filtering procedure, that is, as a method to eliminate UV noise and
permit studying topological excitations:\par
\noindent - It should smooth out the short range fluctuations,
including ``dislocations"\par
\noindent - It should preserve the structure at all physical scales,
including size and location of instantons under any amount of cooling\par
\noindent - It should need no monitoring or any engineering
which would only slow it down without ensuring  stability
and would make the algorithm configuration dependent
(introducing therefore a systematic error).\par
The problem of finding a good cooling algorithm is of course related to that
of finding an action possessing scale invariant classical solutions. From that
point of view perfect actions provide a good basis for such an algorithm. However
the use of many loops and higher representations makes cooling with the
perfect actions computationally rather expensive. Moreover the truncations 
performed  to obtain an action easy to handle
numerically may distort its good scaling properties (the truncated
actions are no longer defined rigorously by a fixed point equation). 
We look for a simpler construction of a good scale invariant action
which does not involve as many loops as  perfect actions and contains
only linear dependence on the links, hence simplifying considerably the
minimization algorithm.  
Since instantons are classical solutions  tree level improvement is
enough to guarantee scale invariance (up to corrections due to replacing
integrals by discrete sums). 
We start therefore with order $a^4$ tree level improvement  
which is analytically defined \cite{MP}, and account for 
higher order corrections by tuning  a free parameter in the action to obtain 
instanton solutions which
are practically scale invariant beyond some small-size threshold. 
This rather straightforward construction leads to an action with very good
scaling properties (which therefore makes it interesting also
as an improved action for Monte Carlo
simulations). It also provides an improved topological charge operator. 
The improved cooling algorithm based on this action and
presented below (see also \cite{MNP}, \cite{MNP96}) fulfills most of the
requirements stated above. The small size threshold is fixed in lattice
units  to $\simeq 2.3 a$ (and therefore shrinks in approaching the continuum
limit). It ensures
that dislocations are eliminated during cooling while physically relevant
instantons are preserved (including their size) if the lattice cut off $a$ 
is chosen small enough. Improved cooling is useful not only for producing good susceptibility
data, but also for studying various conjectured effects of
instantons on the hadronic spectrum or on the chiral transition, by  providing
smooth configurations preserving the large scale structure
of the original ones. 
The algorithm simply minimizes the action in the standard way and involves
no further procedures (heating, blocking, etc).

In the analysis concerning the determination of physical quantities like
topological susceptibility, size distribution, etc  we shall
account for two potential problems, namely \par
\begin{itemize}
\item
 {\it threshold} effects
(for coarse lattices
 $2.3 a$ may already represent a physically relevant distance) and\par
\item
{\it  finite size} effects (small lattices  may not accommodate large,
physical instantons and moreover  for certain sets of  boundary conditions, self-dual $Q=\pm 1$
configurations do not exist in finite volumes \cite{BP}). \par
\end{itemize}
These questions will appear explicitly in our study of
the scaling behavior and of the dependence on size and boundary conditions.
To disentangle threshold from finite size effects we  work at different
values of the lattice spacing, $a$, with the same physical volume.
We also use different lattice geometries (elongated and symmetrical)
with periodic and twisted boundary conditions.
This allows us to study threshold and finite size effects separately,
 and to assert that 
our lattices are such that the physical sizes of interest are affected
by neither of these effects.

 Note that the only structures which in the end will survive cooling are those which are
solutions of the lattice equations of motion. Instantons present in
Monte Carlo configurations are only approximate solutions and  during cooling 
will evolve toward true solutions. Some of the properties of the 
original instanton ensemble change therefore while cooling. 
These concern especially the features of the instanton - anti-instanton (I-A)
pair population (I-A pairs are not minima
of the action, hence they annihilate). Since 
in the physical range of sizes   
practically no other changes in the topological structure occur during cooling 
one can disentangle and study these effects to some extent by following the 
cooling history. Thereby it is very helpful that we can use the improved
charge operator which allows to get meaningful charge density data already after a few 
cooling sweeps. The measured topological charge is an integer within   
${\cal O}( 1{\%})$ already after about 5 sweeps.

The paper is organized as follows:  the improved cooling
is introduced and analyzed in section \ref{s.met}. In section
\ref{s.mc0} we present physical results for the topological susceptibility
as well as for several  properties of the instanton ensemble obtained by
improved cooling on SU(2) configurations generated with the Wilson action.
Both sections contain a number of necessary technical points, which 
however the reader interested only in the physical results may skip.
Section \ref{s.conc} contains a summary of these results and our conclusions.

While our analysis here refers explicitly to $SU(2)$, most relations are valid 
for $SU(N)$.

\section{ The Method of Improved Cooling \label{s.met}}
\subsection{ Improving the Action and the Observables \label{s.def}} 

As was pointed out in Ref. \cite{MP} the stability of instantons
under cooling depends on the sign of the lattice corrections to the continuum action.
On the ground of  pure dimensional arguments  it is easy to see
that deviations from the continuum action for instantons of
size $\tilde{\rho}$ will go as
$ S-8\pi^2 = a_0 \ (a/\tilde{\rho})^2 + {\cal O}((a/\tilde{\rho})^4)$,  
for $\tilde{\rho} >> a$ (from now on we will put a tilde over
dimensionful quantities, otherwise everything will be given in lattice units,
hence $\tilde{\rho}=\rho a$, etc).
This allows to differentiate between ``under-improved"
(e.g., Wilson) actions, for which $a_0 < 0$, and ``over-improved" ones, $a_0>0$; 
under the former instantons shrink, under the latter instantons beyond a 
certain size grow.  This idea was used in \cite{MP} to design
a set of actions, $S(\epsilon)$, based on  a combination of 1$\times$1 and 
2$\times$2 plaquettes, 
with coefficients
\be
c_{1\times 1}= \frac{4-\epsilon}{3},\ \ \ c_{2\times 2}= \frac{\epsilon-1}{48},
\ee
where $\epsilon$ is a free parameter which controls the
sign of the $a^2$ corrections. They are under-improving for $\epsilon > 0$ 
or over-improving for $\epsilon < 0$. Since these actions have been 
thoroughly studied in \cite{MP} we shall not discuss them here any longer.  
They have also been used in \cite{GAM,PK}, to study topology on 
the lattice. 

Following this idea we addressed the problem of designing a lattice
action for which the instanton action deviates as little as
possible from the continuum one. 
Starting from  the results in  \cite{MP} one can construct a one-parameter
set of actions with  no ${\cal O}(a^2)$ and ${\cal O}(a^4)$ corrections
using five fundamental, planar loops of size $m \times n$:               
\bea 
S_{m,n} &=& \ {1 \over {m^2 n^2}} \sum_{x,\mu,\nu} {\rm Tr}
\left( 1 - \ \ \ \ \twooneplaq  \hspace{1.5cm}\right) \nonumber\\
S&=&   \sum_{i=1}^5 c_i \ S_{m_i,n_i}.
\label{e.act1}
\eea
Here  $(m_i,n_i) = (1,1),(2,2),(1,2),(1,3),(3,3)$ for $i=1,\ldots , 5$
(other choices would imply larger loops) and:
\begin{eqnarray}
c_1 &=&(19- 55\  c_5)/9,\ \  c_2 =  (1- 64\  c_5)/9 \nonumber\\
c_3 &=&(-64+ 640\  c_5)/45,\ \  c_4 = 1/5 - 2\  c_5.
\label{e.act2}
\end{eqnarray}
Our choice for an improved action \cite{MNP}, 
\begin{equation}
c_5=1/20,
\label{e.c5}
\ee
is obtained by tuning $c_5$ numerically, taking as criterion that the size of any instanton 
above a certain threshold $\rho_0$ should remain unchanged during 
the cooling procedure (details will be given in section \ref{s.stab}). 
We shall call this loop combination $S(5Li)$, 5-loop-improved.   
We shall also discuss in section \ref{s.stab} two further choices obtained by setting $c_5=0$
and $c_5=1/10$ respectively, which we shall call ``$4Li$" and ``$3Li$".

In addition we use an improved topological charge operator based
on the same set of five loops:
\be
Q =   
\sum_{i=1}^5 c_i  \ Q_{m_i,n_i}. 
\label{q.act1}
\ee
with $c_i$ as in eq.~(\ref{e.act2}) and 
where $Q_{m,n}$ is the naive topological charge operator defined
through loops of size $(m,n)$, i.e.
\be
Q_{m,n} = \ \frac{1}{32 \pi^2}{1 \over {m^2 n^2}} \sum_{x}\sum_{\mu,\nu,\rho,\sigma} \epsilon_{\mu\nu\rho\sigma}
{\rm Tr} \left (\hat{F}_{\mu\nu}(x;m,n)\hat{F}_{\rho\sigma}(x;m,n)
\right )
\ee
with $\hat{F}_{\mu\nu}(x;m,n)$ given in terms
of oriented clover averages of $m\times n$ plaquettes  
\be
\hat{F}_{\mu\nu}(x;m,n) = \frac{1}{8} \ {\rm Im} 
\Biggl\{\ \  \cloverl \hspace{1cm}  + \ \  \cloverr
\Biggr\}.
\ee
\vskip3mm

\noindent It can be easily shown that this definition of the topological
charge is also free of ${\cal O}(a^2)$ and ${\cal O}(a^4)$ corrections.
We shall call 
$Q(5Li)$ the combination for which $c_5$ is set as in eq. (\ref{e.c5}).
The naive, non-improved,  topological charge can be obtained from 
eq. (\ref{q.act1}) by setting $c_1=1$ and $c_{i\ne 1}=0$, we will
call it in what follows $Q(W)$.

Our cooling algorithm exactly minimizes the local action at each 
step. It consists in an iterative replacement of  each link by the
normalized sum of staples connected to it through eq. (\ref{e.act1}).
 It involves no further calibration or engineering. 
 The $Q(5Li)$
charge operator produces reasonable density data already on still rough
configurations and can be used to observe I-A pairs
from the early stages of cooling.

\subsection{ Instanton Description and Size Definitions\label{s.size}}

Before describing how instantons evolve under our improved cooling it is useful
to point out how the instanton size can be determined. This discussion provides
also the basis for our analysis of the size distributions and other
characteristics of the vacuum structure in the next section.  

One can design alternative definitions of the size 
based on the continuum 't Hooft ansatz \cite{bela,tho}. This ansatz
describes an isolated instanton in infinite volume. However, both the finite volume
and the interaction among (anti-)instantons may lead to deviations from this
ansatz (see, e.g. \cite{MP2}). Moreover, the original ``instantons"
produced by Monte Carlo, which are only approximate solutions of the classical 
equations of motion, will typically be deformed by quantum excitations showing 
up in a departure from spherical symmetry. To partially account for these effects 
in identifying these structures we use an 
ellipsoid generalization of the continuum ansatz for an
instanton centered at $\{x_\mu^0\}$:
\begin{equation}
q(x) = {6 N \over { \pi^2 \prod_{\mu=1}^4 \rho_\mu}}
\left[1+\sum_{\mu=1}^4 \left( {{x_\mu - x_\mu^0} \over {\rho_\mu}}\right)^2
\right]^{-4},
\label{e.Qc}
\end{equation}
and similarly for $\hat{s}(x)=s(x)/8\pi^2$ (notice that we have introduced a
normalized action $\hat{S}$, and action density $\hat{s}(x)$ dividing by $8\pi^2$). 
 Here $x_{\mu},\ x_\mu^0$ are taken as real numbers and the
ansatz eq. (\ref{e.Qc}) interpolates between the lattice points. For 
an instanton or anti-instanton the normalization $N$ is +1 or -1, respectively.
The shape of the instanton is characterized concisely by the geometric average size

\be
\rho = \prod_{\mu=1}^4 \rho_\mu^{1/4}.
\label{e.rg}
\ee

\noindent and the quantity:

\begin{equation}
\epsilon = \left[\sum_{\mu=1}^4 \left (\frac{\rho_\mu}{\rho} - 1\right)^2 
\right]^{1/2}
\label{e.ef}
\end{equation}

\noindent introduced as a measure of ``excentricity".

 We also use the partially integrated quantities (``profiles"):

\begin{eqnarray}
s_\mu(x_\mu) &=& \sum_{x_{\nu\neq \mu}} \hat{s}(x) \label{e.st}\\
{\rm or} \ \ q_\mu(x_\mu) &=& \sum_{x_{\nu\neq \mu}} q(x). \label{e.qt}
\end{eqnarray}

\noindent to be fitted by the formula:

\begin{equation}
O_\mu(x_\mu) = {3 N \over { 4  \rho_\mu}}
\left[1+\left( {{x_\mu - x_\mu^0} \over {\rho_\mu}}\right)^{2}
\right]^{-5/2},
\label{e.Qcmu}
\end{equation}

\noindent Here periodicity effects are taken explicitly into account by adding
in the fit satellites $O_\mu(x_\mu \pm N_{\mu})$ ($O=\{q,\hat{s}\}$),
where $N_{\mu}$ is the lattice size.

The size definitions we have used for our analysis are: 
\begin{itemize}
\item[(def.1)]
{\bf$ \rho_{\rp}$}, the so called ``peak" radius, 
\begin{equation}
\rho_{\rp}(Q) =
\left[ \frac{6 }{ \pi^2 |q(x^0)| }\right]^{1/4} 
\label{e.rpk} 
\end{equation}
\noindent with $q(x^0)$ the lattice topological charge density at the center of the
instanton (one could use  the action density $\hat{s}(x^0)$ instead,  
we call the size extracted from it $\rho_\rp(S)$). To obtain the position of the center we interpolate between the lattice points using the ansatz
eq. (\ref{e.Qc}) (see (def.2) below).

\item[(def.2)]
{\bf $\rho_{\rm c}$}, extracted by solving eq. (\ref{e.Qc}) on the central
and nearest neighbor points for $\{x_{\mu}^0, \rho_{\mu}, N\}$ and then taking 
the geometric average eq. (\ref{e.rg}). The corresponding excentricity 
(see eq. (\ref{e.ef})) will be denoted ${\bf \epsilon_c}$. $ \rho_{\rm c}$ and 
$ \rho_{\rp}$ are related through eq. (\ref{e.Qc}):
\begin{equation}
\rho_{\rp} =
\rho_{\rm c}/|N|^{1/4} 
\label{e.rcpk} 
\end{equation}

\item[(def.3)]
{\bf $\rho_{\rm f}$}, obtained with the help of a global fit of the 
configurations to eq. (\ref{e.Qc}). For multi-instanton structures we use a 
linear superposition ansatz. Then we take again the geometric average 
eq. (\ref{e.rg}) to obtain $\rho_{\rm f}$ and we use
eq. (\ref{e.ef}) to obtain the excentricity ${\bf \epsilon_f}$.

\item[(def.4)]
{\bf $\rho^{x_\mu}_\rp$}, based on partially integrated densities eqs. 
(\ref{e.st}, \ref{e.qt}, \ref{e.Qcmu}):
\begin{equation}
\rho^{x_\mu}_\rp (O_\mu) = \frac{3}{4 |O_\mu(x_{\mu}^0)|}
\label{e.rtpk}
\end{equation}
with $O_\mu(x_\mu^0)$ the value  of  $\hat{s},q$
 at the center of the instanton. 
Here we have interpolated between lattice points using
a second order polynomial in $x_\mu$.
We call  ${\bf \rho_{\rm f}^{x_\mu}}$ the size obtained from 
a full fit of the correspondingly integrated ansatz, eq. (\ref{e.Qcmu}).

\item[(def.5)]
{\bf $\rho_{\rm d}(R)$}. For an isolated instanton, and assuming spherical 
symmetry as a first approximation, one can compute  analytically the 
action density integrated over a hyper-sphere of radius $R$  centered at
the instanton position:
\be
\hat{S}_R =  1-\frac{(3R^2+\rho^2)\rho^4}{(\rho^2+R^2)^3}.
\ee\label{previous}
This expression can be used to compute $\rho$ 
given $R$ and $\hat{S}_R$. It will provide a set of definitions of the instanton size
which we call $\rho_{\rm d} (R)$. Of course for an
isolated instanton well described by the continuum ansatz  they should
be $R$ independent.
The way we have proceeded  to extract them
from the data is as follows: locate the center of the instanton, compute the
action contained in a sphere around it of radius $R$ and use
eq. (\ref{previous}) to obtain $\rho_{\rm d}(R)$.
The same can be done using the topological charge instead of the action.
In particular we have used for our analysis $R=\rho_\rp$ and $R=\rho_\rp/\sqrt{2}$. 
\end{itemize}

All these definitions, most of which have already been used
in other works, agree reasonably well with each other as long
as the instanton does not depart too much from the continuum ansatz.
Large instantons, which may be deformed by
finite size effects, or very small instantons, distorted by
lattice artifact effects,  may give rise to differences.
 Of course most of the definitions above are based on a dilute
instanton picture, since they do not take into account the overlap between instantons or
instanton-anti-instanton interactions.  Definition  (def.3) based
on a global fit to a linear superposition of continuum solutions eq. (\ref{e.Qc}),
 can correct for the first of these effects as well as for periodicity
effects, and will ideally account for all the topological
objects present in the configuration.
As a rule, after a few cooling sweeps 
the $5Li$-densities around the peaks of the action are well described by
this ansatz.
We will present below a detailed comparison of the different definitions 
for $\rho$, in connection with the Monte Carlo
results.

Of course, it is possible to use more ``model-independent" definitions for 
the size, e.g. the 4-th root of the volume at whose boundary the density 
has dropped to half the value of the maximum (see, e.g., \cite{MS}). 
However, the physical meaning and the compatibility of such definitions 
among themselves ultimately rely on the continuum ansatz. Since we found 
that the
ellipsoid ansatz can take care of most cases we preferred to stay with definitions based directly on it in order to have less arbitrariness in the 
interpretation of the results and to fully automatize the analysis (with exception of (def.3) all size determinations above can in fact be done {\it on line}).  

\subsection{Tuning of the Action for Optimal Scale Invariance \label{s.stab}}

 To check the effects of tuning the parameter $c_5$
in eq. (\ref{e.act2}), we cooled with different values of $c_5$
a set of instantons of various sizes (generated numerically by cooling 
with $S(\epsilon )$ with various $\epsilon$  in order to vary their sizes),
obtained on $12^4$ lattices with gauge group $SU(2)$. 
Since we want to disentangle small distance from finite
lattice size effects (see next section), we use here twisted 
boundary conditions in the time direction with $k=(1,1,1)$
twist, to ensure that instanton solutions exist on finite lattices
\cite{BP,MP}. In Fig.~\ref{f.iact} we compare the cooling behavior for different
choices of $c_5$ of: (a) the action and (b) the size $\rho_\rp^t$
(def.4)
for the same instanton, as a function of the cooling step.
The two improved actions denoted $3Li$ and $4Li$ correspond
to the settings $c_5=1/10$ and $c_5=0$, respectively, in eqs.~(\ref{e.act1},
~\ref{e.act2}). Although at tree level both of them have no ${\cal O}(a^2)$ 
and  ${\cal O}(a^4)$ corrections they tend 
to act over- or under-improving, respectively, while the choice
$5Li$ preserves the size of all  instantons above the threshold
for an indefinite number of sweeps. This illustrates the 
significance of the tuning eq.~(\ref{e.c5}).
The $3Li$ loop combination, however, is simpler and since it 
also guarantees that physical instantons will not decay under any number of cooling sweeps,
it may be useful for more extensive analyses (large lattices, etc).

Fig.~\ref{f.icoo} shows the behavior of instantons of various initial sizes
under $5Li$-cooling. They remain practically unchanged 
over any practicable number of cooling sweeps provided that 
\begin{equation}
\rho > \rho_0 \simeq 2.3,
\label{e.r0}
\end{equation}
(in units of $a$).
The size of the original instanton has been varied by applying $S(\epsilon)$ 
cooling, therefore the original instantons do not
correspond to minima of $S(5Li)$. Submitting such configurations
to $5Li$ cooling first re-adapts them to the new equations of motion,
implying the small changes which can be observed in the first $\simeq 50$
sweeps of Fig.~\ref{f.icoo}.

In Fig.~\ref{f.ienv} we present a comparison between the size
dependence of the $5Li$-action and Wilson action. 
Instantons of different sizes  are extracted from various stages 
of Wilson cooling of 3 different starting configurations.
The triangles are data from Monte Carlo instantons ($\beta=2.5,\ 12^4$ lattice)
cooled with 300 sweeps of $5Li$ cooling. 
For the improved action  the dependence on the size 
is close to a step-function,  giving  $8\pi^2$ to better than $0.1\%$
for $\rho > \rho_0$ and a steep descent (slightly configuration
dependent) in the interval $(0.8 - 1.0)\rho_0$ \cite{MNP}. 
As Fig.~\ref{f.ienv} shows clearly, there is a small energy barrier 
which prevents the decay of any instanton of size greater than 
$\rho_0$. 

There are simpler actions with no ${\cal O}(a^2)$ corrections, like
the $\epsilon$ action described in section \ref{s.def} with $\epsilon=0$,
which have been shown to improve results on the susceptibility
(see \cite{MIT2} for a QCD study using such an action). Our analysis indicates, however, that
the ${\cal O}(a^4)$ terms may be relatively important in the region of physical sizes 
and they may lead to shifts in the instanton size with cooling, as
already observed by Brower et al \cite{MIT2}. These terms are
 also important  in the threshold region, which tends to loose
its step-function like behavior. The cooling dependence of 
the sizes will also make a scaling test more involved.

\subsection{ Finite Size Effects \label{s.fse}}

 Lattice instantons  may differ from the  continuum ansatz due also
to the finite lattice volume, therefore the question of stability of instantons 
cannot be completely decoupled from finite size effects.
The influence of finite size effects on the quantities
we want to study comes mostly in two ways:
\begin{itemize}
\item[(a)]
They affect the stability of instantons under cooling. Since
on a  periodic hyper-torus self-dual $Q=\pm 1$  solutions
do not exist \cite{BP}, {\it single} instanton configurations will 
always collapse after a while during cooling. 
Of course if the lattice size is much larger than
the instanton size there is a wide window of meta-stability for $Q=\pm1$
which for all practical purposes would amount to stable
instantons. This window can be enlarged by choosing
at least one of the lattice sizes, say $N_t$, very large.
The problem can be completely avoided by 
imposing twisted boundary conditions in one of the lattice directions. In this
case our cooling algorithm guarantees stability of the cooled structures
for any value of the topological charge and an indefinite number of cooling sweeps.
\item[(b)]
There is a  maximal instanton size $\rho_{\rm max}$,  determined  by the
length of the lattice and by the boundary conditions, hence the  large-size 
part of the size distribution will be cut off and distorted.
Due to point (a) above, $\rho_{\rm max}$ will also depend on the total
topological charge of the configuration. We have observed that 
for self-dual $Q=\pm 1$ configurations and 
for symmetric lattices with periodic boundary conditions (p.b.c.) $\rho_{\rm max}^t
\approx N_s/5$ \cite{PK}, for elongated lattices with p.b.c. $\rho_{\rm max}^t
\approx N_s/4$ and for twisted boundary conditions (t.b.c.) $\rho_{\rm max}^t
\approx  2N_s/5$. 
 Moreover large instantons which ``feel" the boundaries will 
typically be deformed by them.
As a consequence for large instantons different size definitions can give quite different
results.
 To obtain a size distribution free
from this effect the  typical physical instanton size has to be less than the 
 maximal instanton size, $\rho_{\rm max}$, allowed by the lattice.
Note however that for higher topological charge $\rho_{\rm max}$
can be larger and may not show such a strong dependence on boundary conditions,
hence the previous estimate is quite conservative.
\end{itemize}  

This provides a general window to retrieve the physical size distribution,
which goes from the cooling threshold $\tilde {\rho}_0=2.3a$ to at
most the maximal instanton size. The threshold is a feature of the 
algorithm and is essential in discarding dislocations; it is fixed by
choosing $\beta$.  As already said $\tilde{\rho}_{\rm max}$ is fixed by the lattice size,
boundary conditions and topological charge of the configuration.

Point (a) is illustrated for $12^4$ and $12^3\times 36$  
lattices on Fig. \ref{f.q1}, where we also show for comparison 
the behavior of a $Q=2$ configuration.  
For $N_s=N_t=12$ p.b.c.  the maximal instanton size (attained by over-improved cooling)
is too close to our stability
threshold  $\rho_0=2.3$, in consequence any isolated instanton
decays quite fast. We expect the region of meta-stability to
be enlarged when $\tilde{\rho}_{\rm max}$ gets away from the threshold (see for
instance the $12^3\times 36$ p.b.c. lattice).
We stress that the instability only affects self-dual $Q=\pm 1$ 
configurations but not multi-instanton configurations and can be completely
avoided by using twisted boundary conditions in at least one lattice direction.
The effect on the susceptibility and size distribution  measured on p.b.c. 
lattices will  be less strong 
for large physical volumes
where the contribution to the susceptibility of configurations with single instantons
is smaller.

 As an illustration of the uncertainties in assigning physical parameters
to very large instantons we present in Table \ref{t.large} data coming from
the analysis of the maximal size instanton on a $12^4$ lattice with twisted 
boundary condition in time. Although this (completely stable) configuration has total charge 1 and action $8\pi^2$ the number of peaks in the energy density is 6
and in consequence a naive assignment of instantons to peaks 
will misleadingly interpret it as 6 different
instantons, with quite large sizes and excentricity. On the other hand an analysis of the time
profile reveals only one peak with much smaller size.

\vskip5mm
\hbox to \hsize{\hfil\vbox{\offinterlineskip
\halign{&\vrule#&\ $#\mystrut$\hfil\ \cr
\noalign{\hrule}
&(x,y,z,t)&&\rho_c &&\rho_p &&\epsilon_c &&N   &\cr
height 4pt&\omit&&\omit&&\omit&&\omit&&\omit&\cr
\noalign{\hrule}
&(3, 4,2, 2) &&7.90 &&7.22 &&0.66 &&-1.43  &\cr
&(4, 6,2, 2) &&7.61 &&7.09 &&0.58 &&-1.33  &\cr
&(1, 1,3, 2) &&7.38 &&6.96 &&0.58 &&-1.26  &\cr
&(6, 7,3, 2) &&7.31 &&7.00 &&0.45 &&-1.19  &\cr
&(10, 4,5,2) &&10.06&&7.52 &&2.21 &&-3.21  &\cr
&(10,10,5,2) &&7.93 &&7.23 &&0.66 &&-1.44  &\cr
\noalign{\hrule}}}\hfil}
\vskip3mm
{{\noindent Table \ref{t.large}: Assignment of parameters made
by the fitting programs to the maximal size instanton ($\hat{S}=1.0000$, $Q=-1.0001$)
on a $12^4$ t.b.c. lattice. $(x,y,z,t)$ represents
the location of the 6 peaks in the action and charge densities.
$N$ is the normalization appearing in the continuum ansatz, eq. (\ref{e.Qc}),
as given by the ``center and nearest neighbors" fit (def.2) in
section \ref{s.size}.
Notice that all the peaks have the same time coordinate, hence
only one peak will be detected in the integrated densities $s_t(t)$
and $q_t(t)$ eq. (\ref{e.Qcmu}) giving a value $\rho_\rp^t=4.66$ (the
time coordinate is selected by the twist).
}\par}
\label{t.large}
\vskip5mm

\section{ SU(2) Topology by Improved Cooling \label{s.mc0}}
\subsection{ Simulation with Wilson Action \label{s.mc}}

The first criterion for the physical relevance of all 
results is how they scale with the removal of the cut-off. 
We work in a region of 
$\beta$ where one expects to see scaling behavior, therefore 
all quantities of physical significance (susceptibility, charge distribution, 
size distribution, etc) should scale according to their dimension 
when approaching the continuum. As
mentioned above, one may expect short-range topological excitations which are unphysical 
in the sense that they do not correspond to attributes of the continuum
physics. These are therefore artifacts of the discretization and will show
up as non-scaling, cut-off dependent structures in the results.
Our algorithm has an inherent cut-off for such structure,
the threshold $\rho_0 \simeq 2.3$. 
Since the threshold is
fixed in lattice units, it shrinks physically and
it should not affect physical sizes if $a$ is
small enough: the physical threshold ${\tilde{\rho}_0} \simeq 2.3a \rightarrow 0 \ {\rm for}\ a \rightarrow 0$.                                 
This we shall test below when  studying the scaling of various quantities.
On the other hand, since Monte Carlo simulations with the Wilson action
copiously produce short range fluctuations one can generally ask whether
this shrinking cut-off at ${\tilde{\rho}_0}$ can indeed ensure the absence of
unphysical fluctuations in cooled configurations when approaching the continuum.
 
Following an argument of Pugh and Teper \cite{PT} we write the
contribution of small instantons of size $\rho_0$ to the partition function as ($N=2$ here)
\be
\left[\rho_0(a)a\right]^{-4}\hbox{e}^{-S_W(\rho_0)/g^2}; \ \   a(\beta) \simeq \Lambda^{-1} \hbox{e}^{-{1 \over {N g^2\beta_1}}},\ \
\beta_1 = {{11N} \over {48\pi^2}}.  \label{e.pt}\nonumber
\ee
The contribution of instantons of a size $\rho$ such that
$\hat{S}_W(\rho)<1/(2\pi^2 N\beta_1)= 24/11N^2 $  diverges for 
$\beta=4/g^2\rightarrow \infty$.
We can calculate $\hat{S}_W(\rho)$ by cooling large
instantons with the Wilson action and measuring the action and the size as the
instantons shrink during cooling (see Fig. 3). This analysis
leads to an UV threshold $1 < \rho_0 < 2$.
It is difficult to get a precise estimate since the dependence of $S_W$ on $\rho$
in this region is sensitive to details of the configuration. We can conclude,
however, that our threshold $\rho_0$ of about 2.3
is fully satisfactory when dealing with configurations produced with the Wilson action:
it ensures removal of the divergent contribution of lattice artifacts,
 without cutting too much from the physical sizes even at rather low $\beta$.

Likewise, the correction due to the finite lattice size should be estimated 
by changing the volume and/or boundary conditions and checking the sensitivity
of the results to these changes. Quite generally, they will
affect structures at the long-range edge of the size distribution.
To estimate the importance of the finite lattice size effects we use both
different physical volumes and boundary conditions.

We base our analysis on the following $SU(2)$ Monte Carlo simulations (heat--bath, 
Wilson action):
\vskip5mm
\hbox to \hsize{\hfil\vbox{\offinterlineskip
\halign{&\vrule#&\ $#\mystrut$\hfil\ \cr
\noalign{\hrule}
& && N_s&&N_t&&\beta && a(\beta)\ (\fm)&&L\ (\fm)&& {\rm b.}\ {\rm cond.}&&\# {\rm conf.}&&\#{\rm sw.}\ {\rm betw.}\ {\rm conf.}&\cr
height 4pt&\omit&&\omit&&\omit&&\omit&&\omit&&\omit&&\omit&&\omit&&\omit&\cr
\noalign{\hrule}
height 4.0pt&\omit&&\omit&&\omit&&\omit&&\omit&&\omit&&\omit&&\omit&&\omit&\cr
&{\bf (1a)} &&12 &&12 &&2.4 &&0.12 &&1.44&&{\rm t.b.c.}&&237&&100 &\cr
&{\bf (1b)} &&12 &&12 &&2.4 &&0.12 &&1.44&&{\rm p.b.c.}&&202&&100  &\cr
&{\bf (1c)} &&12 &&36 &&2.4 &&0.12 &&1.44&&{\rm p.b.c.}&&198&&100  &\cr
&{\bf (2)} &&12 &&12 &&2.5 &&0.085 &&1.02&&{\rm t.b.c.}&&160&&250  &\cr
&{\bf (3)} &&24 &&24 &&2.6 &&0.06 &&1.44&&{\rm t.b.c.}&& 84&&200  &\cr
\noalign{\hrule}}}\hfil}
\vskip3mm
{{\noindent Table \ref{t.dat}: Ensembles of gauge field configurations
analyzed; t.b.c lattices have $k=(1,1,1)$ twist in the time
direction and periodic boundary conditions in space;
p.b.c. lattices have periodic boundary conditions in all directions.}\par}
\label{t.dat}
\vskip5mm
\noindent where we have estimated $a(\beta)$ using the 1-loop phenomenological formula (derived from data in \cite{ukqcd})
\be
a(\beta)= 400 \exp{\biggl \{-\frac{\beta \ln 2 }{0.205}\biggr \}}\ {\rm fm}.
\label{e.ple}
\ee
Here  all  data are taken after 20000 sweeps for thermalization. Part of these data have 
been presented at the Lattice'95 \cite{MNP} and Lattice'96 \cite{MNP96} conferences.

As a rule, during cooling we performed measurements after 5, 20, 50, 150 and 
300 cooling sweeps for lattices (1) and (3), 
and every 20 sweeps for lattice (2) (up to 300).  

We typically calculate the $5Li$ charge and electric and magnetic action 4-dimensional densities 
and thereby the total topological charge and action of the configurations.

\subsection{ Topological Susceptibility and Charge Distribution \label{s.sus}}

 Since the topological charge $Q(5Li)$ stabilizes very
fast to an integer within less than $1\%$, 
we expect 
deviations from the physical values in the susceptibility to be only 
due to {\it threshold} or to {\it finite size} effects. Our results are presented in Fig.~\ref{f.sus}
 and Table~\ref{t.sus}a. 
The susceptibility (from $Q(5Li)$) settles early in the cooling.
 It scales very well with the cut-off (compare the (1) and (3)
data) and shows little dependence on the  
volume (compare lattices (1a-b) and (3) with (1c) and with (2)). Data coming from 
the smaller physical volume seem
to be higher by about 5$\%$, although they are compatible
within errors  with the other two determinations. The effect
can be due to our particular estimation 
of the physical scale eq. (\ref{e.ple}) or it can be a volume effect (that the
lattice size for these data is too small is evident in the size distribution -- see
next section). The small decrease in the susceptibility during cooling 
showing up in the data from lattices (1) ($\beta=2.4$) appears 
to be mostly a {\it threshold} effect. 
Our cutoff on small sizes ($\tilde{\rho}_0=2.3a$) will induce a 
systematic underestimate
in the measurement of the susceptibility since some physical instantons
might be eliminated, but this error goes down with increasing $\beta$ and
can be easily estimated using the size distribution.  As we will see in 
section \ref{s.siz} we obtain size distributions which scale nicely.
Based on the fit to these distributions  
this underestimate should be about $3\%$ for $a=0.12$fm
and less than $1\%$ for $a=0.06$fm.
This is  compatible with the difference between short and long
cooling values (see Table \ref{t.sus}a.) and
 does not change significantly the overall picture. Assuming
the Witten -- Veneziano formula to hold, our result 
$\chi^{1/4} = 200(15)$MeV agrees excellently with
the phenomenological expectation. 

Recent results for $\chi^{1/4}$ 
in SU(2)  \cite{HG}
 are  about  $20\%$ higher (for $\chi$ itself
they are about a factor of two larger).
We believe the discrepancy might be caused by the presence of some dislocations 
in the data of \cite{HG}. 
It would be very interesting to have
data for the  size distribution extracted using inverse blocking. A scaling check
on such distribution will provide an unambiguous way
of deciding whether  dislocations are still present, since their
contribution to the size distribution does not scale.

Further confidence in the correctness  of our results for the topological
susceptibility has been provided by R. Narayanan and P. Vranas \cite{nara}.
They have used the overlap formalism \cite{nara2} to measure the topological
susceptibility on a $12^4$ lattice at $\beta=2.4$ with periodic boundary
conditions (our (1b) lattice). Their results for the topological susceptibility as well as for
the charge distribution are in perfect agreement, within errors, with ours.
 
 Note that the early stabilization of the  susceptibility with cooling
is not only due to the efficiency of the improved
algorithm in removing dislocations, but also to the fact that we use the
improved charge operator $Q(5Li)$.
 To illustrate  this we  present in Table~\ref{t.wil}b
the values of the topological susceptibility, for lattices (1a)
and (2) in Table \ref{t.dat}, extracted from  the measurement of
$Q(W)$  over the same set of $5Li$ cooled configurations.
\hbox to \hsize{\hfil\vbox{\offinterlineskip
\halign{&\vrule#&\ $#\mystrut$\hfil\ \cr
\noalign{\hrule}
&&&(1{\rm a)} &&(1{\rm b)}&&(1{\rm c)}  &&(3)  && &&(2)&\cr
height 4pt&\omit&&\omit&&\omit&&\omit&&\omit&&\omit&&\omit&\cr
\noalign{\hrule}
&{\rm sw}&&5Li&& 5Li&&5Li &&5Li&&{\rm sw}&& 5Li&\cr
height 4pt&\omit&&\omit&&\omit&&\omit&&\omit&&\omit&&\omit&\cr
\noalign{\hrule}
height 4pt&\omit&&\omit&&\omit&&\omit&&\omit&&\omit&&\omit&\cr
&0  &&161(5)&&     &&       &&          &&0  &&219(8) &\cr
&5  &&199(3)&&200(5) &&201(5) &&205(16) &&   &&       &\cr
&20 &&200(3)&&200(6) &&199(5) &&199(15) &&20 &&212(7) &\cr
&50 &&198(3)&&198(6) &&195(5) &&197(15) &&60 &&212(7) &\cr
&150&&196(4)&&195(6) &&192(6) &&197(15) &&140&&212(7) &\cr
&300&&195(4)&&193(6) &&189(6) &&194(14) &&300&&211(7) &\cr
\noalign{\hrule}}}\hfil}
\vskip3mm
{{\noindent Table \ref{t.sus}a: $^4\hspace{-1mm}\sqrt{\chi}$ (MeV) 
computed from the improved charge operator $Q(5Li)$,
{\it vs} number of cooling sweeps.
Errors have been estimated using the jackknife method.}\par}
\label{t.sus}
\vskip5mm
\noindent The susceptibility stabilizes
much earlier with cooling if the improved operator is used. Indeed
even when no cooling at all is performed, the deviation from the asymptotic
value is at most $20\%$ for $Q(5Li)$ while it is twice as much for
the naive charge operator (after 5 sweeps $Q(5Li)$ is already asymptotic
while $Q(W)$ is still $10\%$ off). The results obtained
with $Q(5Li)$ after only a few cooling sweeps deviate by no more than $0.5\%$ 
from what is  obtained
by approximating the charge by the closest integer. Even without cooling
the deviation is only about $1-2\%$. 
This rounding procedure also works for $Q(W)$, but only after a long cooling:
approximating the charge by the closest integer  after 300 sweeps we  
obtain $^4\hspace{-1mm}\sqrt{\chi}= 194(4)$, 212(7) (MeV)
for (1a) and (2) respectively, in perfect agreement  with the results from $Q(5Li)$.
The importance of improving  the topological charge operator 
for non-smooth configurations has been recently stressed in~\cite{dg3}.

\vskip5mm
\hbox to \hsize{\hfil\vbox{\offinterlineskip
\halign{&\vrule#&\ $#\mystrut$\hfil\ \cr
\noalign{\hrule}
&&&(1{\rm a)} && &&(2)&\cr
height 4pt&\omit&&\omit&&\omit&&\omit&\cr
\noalign{\hrule}
&{\rm sw}&&{\rm Wilson}&&{\rm sw}&&{\rm  Wilson}&\cr
height 4pt&\omit&&\omit&&\omit&&\omit&\cr
\noalign{\hrule}
height 4pt&\omit&&\omit&&\omit&&\omit&\cr
&0  &&123(3)&&0 &&167(6) &\cr
&5  &&179(3)&&  &&       &\cr
&20 &&186(3)&&20 &&201(6) &\cr
&50 &&188(3)&&60 &&204(7) &\cr
&150&&188(4)&&140&&204(6) &\cr
&300&&187(4)&&300&&204(6) &\cr
\noalign{\hrule}}}\hfil}
\vskip3mm
{{\noindent Table \ref{t.wil}b: $^4\hspace{-1mm}\sqrt{\chi}$ (MeV)  
computed from the naive charge operator $Q(W)$
{\it vs} number of cooling sweeps.
\par}}
\label{t.wil}
\vskip5mm

 The topological charge distribution is presented in Fig.~\ref{f.ch} for
various boundary conditions and numbers of cooling sweeps. The differences
induced by the former are related mostly to the instability of self-dual charge $\pm 1$
configurations for periodic boundary conditions (the p.b.c. data show typically
more configurations at $Q=0$ and fewer at $Q=\pm 1$ than the t.b.c. data, but the statistics is not sufficient for a significant
  signal). The disappearance of
 some narrow peaks (beyond pair annihilation) when going from 20 to 300 cooling sweeps does not lead to significant changes
in the distribution.
The shape of the charge distribution
agrees well with a gaussian of width given by the susceptibility. These data are also nicely corroborated by those 
from \cite{nara}.

\subsection{ Instanton Shape and Size Distribution \label{s.siz}} 

For a systematic analysis of the local structure we developed a number of programs to automatically
recognize and evaluate the local features. We think it useful to give some technical details about the procedure, to permit an overview
of the various constraints and uncertainties. 

Essentially these programs proceed by
finding first the peaks in the action and in the topological charge density. 
The identification of the peaks proceeds along the following three steps:
\begin{itemize}
\item
Look for peaks in the electric and magnetic parts of the action density,
$E^2(x)$ and $B^2(x)$.
Assign a peak to every lattice point whose  density is maximal with respect to its
nearest neighbors. We also assign a normalization $\pm 1$ to each peak depending on
whether its topological charge is positive or negative.
\item
Apply a cut that removes from the set of peaks those which are so flat that
$\rho_{\rp}(S)$  is larger than $N_s$.
Typically small fluctuations, not associated to instantons, or
remnants of I-A annihilation will give rise to very low peaks
in the energy density, which misleadingly translate via the $\rho_\rp$ formula
eq. (\ref{e.rpk})
into very large sizes. With this cut we suppress part of these effects.
\item
Apply a  cut from self-duality by comparing the locations of peaks in  $E^2(x)$
and $B^2(x)$ (with equal normalization). Only peaks for which the distance between locations
is less than or equal to $2a$ are kept.
\end{itemize}
After performing these cuts we have observed that there is an abundance of close peaks
separated from each other no more than 1 lattice spacing along each direction.
This behavior has been also pointed out in \cite{MS}. We have adopted the attitude
of identifying such peaks as being only one. We believe  that the effect is mostly due to locating
peaks by looking only at nearest neighbors and not at points separated along
the diagonal (slightly excentric instantons located diagonally between the lattice
points may lead to such a signal). 
The effect of this last cut  reduces the number of peaks by 
about $8\%$ independently of the number of cooling sweeps.
It is important to point out that the cut is applied in lattice
units so if there is indeed a physical effect associated to
close peaks it should be possible to detect it anyway by reducing the
lattice spacing.
In section  \ref{s.dis} we will discuss further this effect
by comparing results from lattices with $a=0.12$ fm and $a=0.06$ fm.

The typical level of accuracy achieved by these programs turned out to be above
$90\%$, the failures being mostly related to structures affected by threshold or by
finite size effects (e.g., wrong normalization, misinterpretation of a wide,
deformed instanton as a multiple peak structure, etc). While global quantities like total charge or
topological susceptibility are not affected by such uncertainties, the latter will be reflected
in systematic errors concerning, e.g., the small and large size edges of the
size distributions. The errors we quote in this paper are statistical only. 
But we now proceed to estimate the systematic errors. 

\subsubsection{ Determination of Instanton Shape and Size  \label{s.shap}}

Both finite size effects and instanton - (anti-)instanton interaction
may deform the spherically symmetric, continuum ansatz and  we did not want to bias a priori the information about instanton shape by 
assuming spherical symmetry.
We decided therefore to evaluate  simultaneously several of the
size definitions in section \ref{s.size}. A comparison between them and more specifically
a measurement of the excentricity (see eq. (\ref{e.ef})) provides 
a good 
estimate of the departure from the continuum ansatz.
Our analysis proceeds along two directions:
\begin{itemize}
\item[(a)]
Extracting ``on line"  sizes
corresponding to definitions 1, 2, 5 in section
\ref{s.size} (i.e. $\rho_\rp$, $\rho_c$ and $\rho_d(R)$ for $R=\rho_\rp$
and $R=\rho_\rp/\sqrt{2}$). We also extract the excentricity $\epsilon_c$
using (def.2)  in section \ref{s.size}.
\item[(b)]
Performing by MINUIT global fits of the $5Li$ charge
and action densities using eq.~(\ref{e.Qc}). In most cases this ansatz
 fits  rather well the $5Li$-densities
(charge and action) of isolated instantons already after a few
cooling sweeps. The simple superposition extension of this ansatz can 
also handle overlapping structures in multi-peak fits over clusters. 
\end{itemize}
We tried to assess the uncertainty of the analysis self-consistently
by comparing all the size definitions.

The global fits (b) start with initial information about
the location and size of the peaks provided by (a).
Clusters are first identified, by
considering as neighbors peaks at a distance less than their average
radius. For this we use the size estimate from (a),
or a fraction thereof if we
want to limit the maximal cluster size (MINUIT would not cope with more than 5 peaks at a time). Once identified, the clusters 
are fitted separately with a superposition ansatz. The MINUIT fit uses
8 or 9 parameters per peak (in the latter case the normalization  $N$ in eq.~(\ref{e.Qc})
is taken as a real number and permitted
to vary). The fitting procedure tends to fail in the cases where also
the peak assignment is not well defined (see the discussion
in sect. \ref{s.fse}). For the description of the topological structure uncovered in this way we shall use the geometric average size eq. (\ref{e.rg}) 
and the ``excentricity" eq. (\ref{e.ef}). Since the global fits are rather
expensive we performed them only on subsets of 10 configurations for the 
(1a) and (1b) lattices and 5 configurations on (1c) (for the $24^4$ lattices we
could not perform global fits on our workstations), therefore the data for
$\rho_f$ and $\epsilon_f$ ((def.3), section \ref{s.size}) have large statistical errors. 
 
A comparison between the fit analysis, $\rho_{\rp}$,  (def.1)
and  $\rho_c$ (def.2) is given
in Table \ref{t.ex}a. 
 The differences in the results for size  and the excentricities
are presented as the root mean squares 
$\delta \rho/\rho|_{\rm rms}$ and $\epsilon_f|_{\rm rms}$, with
$\delta \rho/\rho = 2(\rho_{\rp}-\rho)/(\rho_{\rp}+\rho)$. Here 
 $\rho = \rho_c$ or $\rho = \rho_f$ as obtained from the following fits:
``f4" (free norm cluster fit with up to 4 peaks in a cluster),
 ``fn" (cluster fit with up to 4 peaks in a cluster with the 
norm $N=\pm 1$ as assigned by (a)) and ``fs" 
(single peak fit with the norm $N$ taken as free parameter).
The figures in brackets correspond to discarding peaks for which 
one or more of the sizes are larger than the lattice size -- these cases typically prove to be I-A pairs in the course of annihilation or simply
failures of the fit program in the presence of very distorted structures.
They affect $5 - 10\%$ of the peaks.

As can be seen, the fitted radius averaged over directions agrees
quite well with $\rho_{\rp}$  (def.1). 
From the data in Table \ref{t.ex}a we estimate a total error of less
than $10\%$ for long cooling: we reproduce this (informal)
estimation  $\sigma(\rho)$ in Table \ref{t.ex}b.
Both $\sigma(\rho)$ and $\epsilon_f|_{\rm rms}$ decrease with
the number of cooling sweeps, which is understandable as an effect of smoothing
and of the disappearance of I-A pairs.

In Fig.~\ref{f.rfp} we compare the size distributions obtained
from the $\rho_{peak}$ and $\rho_c$ data. We also show here the ``excentricity" distribution $\epsilon_c$. As is evident
in the cross-correlation between                                       
size and excentricity on Fig.~\ref{f.rfp}, larger instantons depart more from
the spherical shape. Also, the average excentricity decreases with cooling.
The excentricity can have at least four origins:

\begin{itemize}
\item
sensitivity of the fit procedure to details of the configuration not 
considered in the ansatz, e.g., the presence of nearby instantons;
\item
interaction among (anti-)instantons, which is not taken care of by the ansatz;
\item
physical fluctuations (e.g., higher momentum states) in the ensemble of instantons produced in the Monte Carlo simulation at physical sizes;
\item 
distortion of the true lattice instantons at large sizes due to finite volume effects.
\end{itemize}
Both the first and second effect are stronger if the instantons have a large overlap, 
therefore they should decrease with cooling as a result of the dilution of 
the ensemble due to pair annihilation. For the same reason they also could affect 
large instantons more than small ones. The smoothing out of physical fluctuations
during cooling will tend to decrease the average excentricity for instantons in the 
range of sizes not affected by finite volume effects, since there the  exact 
solution is spherically symmetric; this does not necessarily hold for large instantons, which may
be very deformed asymptotically.
Fig.~\ref{f.rfp} shows that excentric instantons are rare.
We have explicitly verified that the 4 axes $\rho_{\mu}$ of our ellipsoidal
fits are strongly correlated.
Departure from spherical
symmetry is moderate, and well accounted for by our ellipsoidal ansatz,
eq.(\ref{e.Qc}).

 In  Fig.~\ref{f.rqs}  we compare sizes obtained from definitions 1 and 5
(see section~\ref{s.size}): $\rho_\rp(S)$, $\rho_\rp(Q)$ or $\rho_d(R)$.
All should agree if the continuum ansatz is correct. As we can see, the differences are
typically well within the uncertainty estimated on the basis of the fits,
and mostly affect large instantons of size above $N_s/2$.
The effects related to large instantons are clearly reflected in Fig.~\ref{f.b25} where we compare
size distributions after 300 cooling sweeps extracted from $\rho_{\rp}(Q)$
(def.1) and $\rho_{\rm f}^t$ (def.4) for lattice (2) 
($12^4$ at $\beta=2.5$, physical size approximately 1fm).
 The $\rho_{\rm f}^t$ distribution has a strong large size cutoff around $N_s/2$
while the $\rho_{\rp}(Q)$ one is quite biased towards
large sizes. 
In addition the number of peaks in the energy density is larger
by as much as $40\%$  than the one extracted from spatially integrated densities 
(it decreases from 98 to 61 after integration for 70 configurations analyzed).

 A further measurement of the good agreement between different size definitions
is provided by the value of the normalization entering in the
density formula eq. (8) (see also eq. (15)). In Fig. \ref{f.norm}
we present the  histogram for the distribution of normalizations
obtained from the ``center and nearest neighbors" analysis (def.2). It is
nicely centered about $\pm 1$ with a dispersion that decreases with
the number of cooling sweeps due to effects analogous to the ones quoted
above in relation with the excentricity.

\vskip1.5cm
\hbox to \hsize{\hfil\vbox{\offinterlineskip
\halign{&\vrule#&\ $#\mystrut$\hfil\ \cr
\noalign{\hrule}
& && && \hskip5mm (1 {\rm a})\ \ \ {\rm t.b.c} &&\hskip5mm (1 {\rm b}) 
\ \ \ {\rm p.b.c.}&&\hskip5mm  (1 {\rm c})\ \ \ {\rm p.b.c.\ elongated}&\cr
height 4pt&\omit&&\omit&&\omit&&\omit&&\omit&\cr
&{\rm sw}\!&&{\rm fit} &&\# {\rm peaks}\hskip2mm\frac{\delta \rho}{\rho}|_{\rm rms}\hskip.3cm 
\  \epsilon_f|_{\rm rms}&&\# {\rm peaks}\hskip5mm\frac{\delta \rho}{\rho}|_{\rm rms}\hskip.3cm 
\  \epsilon_f|_{\rm rms}&&\# {\rm peaks}\hskip5mm\frac{\delta \rho}{\rho}|_{\rm rms}\hskip.3cm
\  \epsilon_f|_{\rm rms}&\cr
height 4pt&\omit&&\omit&&\omit&&\omit&&\omit&\cr
\noalign{\hrule}
height 4pt&\omit&&\omit&&\omit&&\omit&&\omit&\cr
&&&      c&&\ 98\ \ \ \ \ \ \  0.125 \hskip.6cm 0.427 &&
\ 103\ \ \ \ \ \ \  0.117 \hskip.6cm 0.333&&
\ 172\ \ \ \ \ \ \  0.079 \hskip.6cm 0.552 &\cr
&&&      &&(96)\ \  \ \ (0.115)\hskip3mm (0.318) &&
 (101)\ \  \ \ (0.109)\hskip3mm (0.290)  &&
 (171)\ \  \ \ (0.076)\hskip3mm (0.552)  &\cr
&&&      fn&&\ 98\ \ \ \ \ \ \   0.125 \hskip.6cm 1.427 &&
\ 103\ \ \ \ \ \ \  0.203 \hskip.6cm 1.563&&
\ 172\ \ \ \ \ \ \  0.099 \hskip.6cm 0.622 &\cr
&20\! &&      &&(96)\ \ \ \  (0.123)\hskip3mm (0.386) &&
\ (94)\ \ \ \ \ (0.205)\hskip3mm (0.375)  &&
 (168)\ \ \ \  (0.093)\hskip3mm (0.414)  &\cr
&&&     f4&&\ 98\ \ \ \ \ \ \  0.176 \hskip6mm  0.536 &&
\ 103\ \ \ \ \ \ \  0.191 \hskip.6cm 0.962 &&
\ 172\ \ \ \ \ \ \  0.184 \hskip.6cm 0.808 &\cr
&&&      && (95)\ \ \ \  (0.170)\hskip3mm (0.326) &&
 (101)\ \ \ \  (0.158)\hskip3mm (0.308)  &&
 (165)\ \ \ \  (0.176)\hskip3mm (0.404)  &\cr
&&&     fs&&\ 98\ \ \  \ \ \ \  0.140 \hskip.6cm 0.527 &&
\ 103\ \ \ \ \ \ \     0.157 \hskip.6cm 0.352   &&
\ 172\ \ \ \ \ \  \     0.144 \hskip.6cm 0.384 &\cr
&&&      && (97)\ \ \ \  (0.136)\hskip3mm (0.293) &&
 (101)\ \ \  \  (0.158)\hskip3mm (0.308)  &&
 (171)\ \ \ \  (0.143)\hskip3mm (0.375)  &\cr
height 4pt&\omit&&\omit&&\omit&&\omit&&\omit&\cr
\noalign{\hrule}
height 4pt&\omit&&\omit&&\omit&&\omit&&\omit&\cr
&&&      c&&\ 17\ \ \ \ \ \ \   0.033 \hskip.6cm 0.211 &&
\ \ 15\ \ \ \ \ \ \  0.246 \hskip.6cm 0.248&&
\ \ 28\ \ \ \ \ \  \ 0.065 \hskip.6cm 0.221 &\cr
&&&      && (17) &&
\ (13)\ \ \ \ (0.084) \hskip.35cm (0.115) &&\ (28)
   &\cr
&&&      fn&& \ 17\ \ \ \ \ \ \  0.095 \hskip.6cm 0.295 &&
\ \ 15\ \ \ \ \ \ \ 0.222 \hskip.6cm 0.253&&
\ \ 28\ \ \ \ \ \ \  0.100 \hskip.6cm 0.409 &\cr
&300\! &&      && (17) &&
\ (13)\ \ \ \ (0.083) \hskip.35cm (0.153) &&\ (27)\
 \ \ \ \ (0.060) \hskip.25cm (0.211)   &\cr
&&&     f4&&\ 17\ \ \ \ \ \ \  0.101 \hskip6mm  0.275 &&
\  \ 15\ \ \ \ \ \ \ 0.179 \hskip.6cm 0.274 &&
 \  \ 28\ \ \ \ \ \ \ 0.064 \hskip.6cm 0.246 &\cr
&&&      && (17) &&
\ (13)\ \ \ \ (0.021) \hskip.35cm (0.127) &&\ (28)
   &\cr
&&&     fs&&\  17\ \ \ \ \ \ \ 0.101 \hskip.6cm 0.272 &&
\  \ 15 \ \ \ \ \ \ \    0.238 \hskip.6cm 0.391   &&
\ \ 28 \ \ \ \ \  \  \  0.092 \hskip.6cm 0.274 &\cr
&&&      && (17) &&
\ (13)\ \ \ \ (0.021) \hskip.35cm (0.121) &&\ (28)
   &\cr
height 4pt&\omit&&\omit&&\omit&&\omit&&\omit&\cr
\noalign{\hrule}
\noalign{\hrule}}}\hfil}
\vskip3mm
{{\noindent Table \ref{t.ex}a: Comparison of fit sizes and excentricities }\par}
\label{t.ex}
\vskip1cm

\vskip5mm
\hbox to \hsize{\hfil\vbox{\offinterlineskip
\halign{&\vrule#&\ $#\mystrut$\hfil\ \cr
\noalign{\hrule}
& &&   (1 {\rm a}) && (1 {\rm b}) && (1 {\rm c})&\cr
&{\rm sw}&& \sigma(\rho)&&\sigma(\rho)&&\sigma(\rho)&\cr
height 4pt&\omit&&\omit&&\omit&&\omit&\cr
\noalign{\hrule}
&20        &&0.14\ (0.12) &&0.16\ (0.15)&&0.14\ (0.14) &\cr
&300       &&0.06 \ (0.06) &&0.20\ (0.05) &&0.08\ (0.06) &\cr
\noalign{\hrule}
\noalign{\hrule}}}\hfil}
\vskip3mm
{{\noindent Table \ref{t.ex}b: Informal estimation of systematic errors}\par}
\vskip1cm

A general conclusion from the analysis described above is the good agreement,
up to at most 10$\%$ discrepancy, of all the size definitions.
Discrepancies appear due mostly to finite size effects and I-A pairs in the late stages of
annihilation, which typically reflect in a departure
from spherical symmetry 
and a wrong assignment of peaks.
Based on these observations, we decided to quote in the following
results for $\rho_{\rp}$ (def.1, eq.~(\ref{e.rpk})) only.
 (In \cite{MNP} the instanton sizes for lattice (2) were
extracted  from the t-profiles -- see eq.~(\ref{e.rtpk})).
Systematic errors in the determination of the size are estimated
to be at most $10\%$.

For the lattice (3), we used lattice coordinates in eq.~(\ref{e.rpk})
to define the instanton center and no
continuous interpolation of  $O_\mu(x_{\mu})$ in ${x}_\mu$ was attempted.
This may lead to an overestimation of the instanton size by up to
$1/\rho^2$ which for the smallest stable instantons allowed by
our algorithm, $\rho=2.3$, amounts to at most  about 15$\%$, while for
the typical size instanton it is below $2\%$.

\subsubsection{ Instanton Size Distribution \label{s.sizd}}

The results for the size distributions
are presented in Fig.~\ref{f.siz}. We only compare
the size distributions for lattices (1) and (3) since for lattice (2)
the large-size edge of the distribution is
strongly affected by finite size effects (see discussion
in section \ref{s.shap} and Fig. \ref{f.b25}).
$P(\tilde{\rho})$ counts the number of instantons at distance 
$\tilde{\rho}$ normalized in such a way that the integral 
over $\tilde{\rho}$ is equal to 1.

The primary criterion for the physical significance of these distributions is
how they scale in approaching the continuum limit. As we see from Fig.~\ref{f.siz}
the
threshold (indicated by the vertical dashed line) is gradually removed
with decreasing $a$  and the distributions scale
nicely above it. The data at $\beta=2.6$ show that most of the
physically relevant structure is in a well defined region above the threshold,
where the latter has no influence. Comparison of the (1) and (3) data
shows that at $\beta = 2.4$ the bulk of the size distribution already begins
to get out of the threshold region. The slight decay of the susceptibility
with cooling
observed in Table \ref{t.sus}a and Fig. \ref{f.sus} for $\beta=2.4$ can be completely
understood as due to the disappearance of a few, small but still
physically relevant instantons close to the threshold (compare
in Fig.~\ref{f.siz} the distribution of instantons close to the
threshold between 20 and 300 cooling sweeps).
 This good scaling of the distributions
when changing the lattice cut off by a factor 2  gives evidence for
their physical relevance.
The stability under cooling of the distributions 
is also a check of the physical relevance of the results and the
success of our algorithm in preserving the structure of the original configurations.

In Table~\ref{t.spois} we give the data for the non-normalized size distributions,
the value $\tilde{\rho}_M$ of the instanton size at the peak of the distribution
 and the ``width" $\tilde{w}$ of the 
distribution extracted from a fit
\be
P(\tilde{\rho}) \propto \ \tilde{\rho}^{\ 7/3} \exp \left (-\frac{\tilde{\rho}^{\ p}}
{\tilde{w}^{p}} \right )
\label{e.rfit}
\ee
which for small $\rho$ reproduces the semi-classical result
\be
P(\tilde{\rho})\propto  \tilde{\rho}^{7/3} 
\ee
and for large sizes introduces a damping  which cuts the infrared
divergence. This ansatz can be derived by using the semi-classical
distribution with a modified running coupling constant which stops
running beyond some value of $\tilde{\rho}$ (see \cite{models}).
Other plausible ans\"atze, like a Gaussian
distribution, or $\rho^{7/3}$ times a Gaussian, give significantly 
poorer fits.

The curve in Fig.~\ref{f.siz}  represents a fit with
eq.~(\ref{e.rfit}) to the average of all (1) data after
20 cooling sweeps and the (3) data after 50 cooling sweeps 
(leaving out the data below threshold for (1))
with a $\chi^2$ per degree of freedom of 0.9.
This fit gives a typical 
instanton size
 $\tilde{\rho}_M=0.43(5)$ fm,  a 
full ``width" $\tilde{w}= 0.47(9)$ fm and a value $p=3(1)$ for the exponent.

If the action used in the Monte Carlo simulation introduces spurious structure {\it above} the threshold, this
structure will of course be seen by our algorithm, and should be 
identified as such by a scaling analysis. Since the Wilson action $\hat{S}(W)$ for instantons
of size $\rho_0$ is lower than 1 (see Fig. \ref{f.ienv}) we may expect it to
produce in the Monte Carlo simulation an unwanted 
excess of instantons there. This 
may lead to an ``artificial"
increase in the size distribution in the region just above $\rho_0$.
This effect should disappear when the physical sizes involved move away 
from the threshold. From Fig.~\ref{f.siz} we see that the relevant part
of the size distribution scales. 
This is indication that the effect under discussion is small,
and that the size distribution we measure may be considered physical.

\vskip4mm
\hbox to \hsize{\hfil\vbox{\offinterlineskip
\halign{&\vrule#&\ $#\mystrut$\hfil\ \cr
\noalign{\hrule}
$\ \tilde{\rho}$&(\fm)&&\!0.18 \! &&\!0.30\!  &&\!0.42\!  && \!0.54\!
&&\! 0.66\!  &&\! 0.78\!   &&\! 0.90\!  &&\! 1.02\!  &&\! 1.14\!  &&\! 1.26\! 
&&\! 1.38\!  &&\!&&\ \!\tilde{\rho}_M\!&& \! \chi^2/dof\!&\cr
height 3pt&\omit&&\omit  &&\omit &&\omit  &&\omit &&\omit
&&\omit &&\omit &&\omit  &&\omit &&\omit &&\omit &&\omit&\cr
\noalign{\hrule}
{\bf (1a)}&20{\rm sw}\! &&0.55 &&2.50 &&3.24 &&2.80 &&1.14 &&0.34 &&0.01  &&0.00 &&0.00
&&0.00 &&0.00 &&\! &&0.43 &&1.0&\cr
&300{\rm sw}\!&&0.00 &&0.36 &&0.48 && 0.40 &&0.37 &&0.21 &&0.17 &&0.04 &&0.04 
&&0.03&&0.01&&\!&&0.44  && 0.6&\cr
\noalign{\hrule}
{\bf (1b)}&20{\rm sw}\!&&0.60 &&2.46 &&3.37 &&2.72&&1.18&&0.23&&0.01&&0.00 &&0.00 
&&0.00&&0.00&&\!&&0.43 &&0.3 &\cr
&300{\rm sw}     \!&&0.03 &&0.47&&0.68  &&0.31 &&0.16&&0.07 &&0.01 &&0.01 
&&0.01 &&0.00 &&0.00&&\!&&0.38  &&1.1 &\cr
\noalign{\hrule}
{\bf (1c)}&20{\rm sw}\!&&1.82 &&7.85\! &&9.48\! &&7.87\! &&3.70&&0.80&&0.05
&&0.00&&0.00  &&0.00 &&0.00 &&\!&&0.43   &&1.6 &\cr
&300{\rm sw}\!&&0.07 &&0.73&&1.18&&1.16&&0.70&&0.29&&0.17&&0.07&&0.06&&0.04&&0.04&&\!&&0.46  &&3.&\cr
\noalign{\hrule}
{\bf (3)}&50{\rm sw} \!&&2.36 &&3.26&&4.32&&3.65&&1.58&&0.44&&0.08 &&0.02 &&0.01 &&0.00&&0.00&&\!&&0.39  &&2.5 &\cr
&  300{\rm sw}\!&&0.57&&0.98 &&0.82 &&0.76 &&0.40 &&0.25 &&0.12 &&0.02 &&0.01&&0.00&&0.00&&\!&& 0.37
  &&0.5 &\cr
\noalign{\hrule}
\noalign{\hrule}}}\hfil}
\vskip1mm
{{\noindent Table \ref{t.spois}: Non-normalized size distributions and typical 
instanton size
(we only divide by the number of configurations;
notice that we have changed the binning for the $24^4$ lattice with
respect to Fig.~\ref{f.siz}).
For the bins $\tilde{\rho}=0.06$fm and below there are no contributions.
}\par}
\label{t.spois}
\vskip5mm

 The typical instanton size we obtain can be compared with previous results
obtained in Refs. \cite{PV} and more recently \cite{MS} and \cite{MIT1,MIT2}. 
Polikarpov and Veselov \cite{PV} cool using the Wilson action for a
small number of sweeps  and quote an average size of about 0.38 fm.
 Michael and Spencer \cite{MS} 
use a cooling algorithm based on the Wilson action and adjusted
to provide a meta-stability window for the instantons.
They have results
on $16^4$ and $24^4$ lattices at  $\beta=2.4$ and $\beta=2.5$ respectively.
The peak of their size distribution depends on the lattice size, obtaining
$\tilde{\rho}_M=0.38$ fm and 0.26 fm for $16^4$ and $24^4$
respectively. We believe that their method is subject to systematic errors
induced by the unavoidable shrinking of instantons
under  Wilson cooling. They have carefully adjusted their algorithm
to obtain the same ``degree" of cooling on both lattices but still
the difference is large.
Our method provides a safe way of ensuring no distortion of physical
structures without need for any further tuning or calibration of
the algorithm.
Chu et al \cite{MIT1} use a convolution with the continuum ansatz
of the density correlations
obtained by cooling with Wilson action for the SU(3) theory. They quote 
an instanton size of about $0.36$ fm after 20 cooling sweeps.
Brower et al \cite{MIT2} use an ${\cal O}(a^2)$ improved action in connection with a relaxation variant of cooling for $SU(3)$ theory without and with
dynamical fermions. They apply an instanton identification procedure also
based on the continuum ansatz
similar to the method used in \cite{MNP96} and further developed here. Since the maximum in the size distributions shown in 
\cite{MIT2} drifts from about 0.4 fm after 20 ``cooling" sweeps to 
about 0.5 fm after 50 sweeps (presumably as a result of the ${\cal O}(a^4)$ terms
in the action) the comparison is however difficult.

Finite size effects can be estimated by comparing (1a,b,c). The results
agree perfectly for short cooling.
After 300 cooling sweeps there are several sources of systematic errors which can give rise
to slight differences.
For lattices (1a,b,c), $a=0.12$fm, and
there are some physically relevant instantons
close to the threshold which disappear with cooling.
This fact tends to induce a small overall shift of the (1) size
distributions towards larger sizes. This is  observed
for the (1a) and (1c) data (see Table \ref{t.spois}), however for 
(1b) ($12^4$, p.b.c) there is instead a shift towards smaller sizes
(see Table \ref{t.spois} and the increase of the number
or instantons of size $\sim 0.4$ fm indicated on Fig. \ref{f.fse}
by an arrow). 
We believe this effect to be
mostly due to the instability of isolated instantons with p.b.c.
(see section \ref{s.fse}). To check this conjecture we left out of the ensembles
(1a,b,c) configurations with action $S=8\pi^2$. 
In Fig. \ref{f.fse} we compare the original distributions
(1a,b,c)  with the ones generated from them by removing
$S=8\pi^2$ configurations (keeping the normalization
of the original distributions), which we denote by (1a',b',c')
respectively.
We  see
that the boundary conditions dependence is reduced, supporting the
conclusion that for $L=1.44$fm 
this is the dominant finite size effect.

We can now go back to the discussion in section \ref{s.fse} about
the strong finite size effects on the data (2) taken at $L=1.02\ \fm$.
As we have just discussed, physically relevant instantons have a
typical size of 0.43 fm. The bulk of the size distribution is thus
concentrated in the region   where we  have argued the effect of
the boundary begins to be strong and distort the instantons.
$L=1.44\ \fm$ however seems to be just safe enough.
 A general conclusion we extract from this analysis 
is that to be free of finite size effects the lattice
physical size should be at least 3 times the typical physical instanton size.

The cutoff at $\tilde{\rho} \approx 2.3 a$ of our method eliminates lattice
artifacts by construction. Another approach would consist in measuring
the size distribution of {\em all} objects, physical and unphysical,
then to separate the ones from the others by a scaling study.
This is the approach attempted in \cite{PK}, using ``over-improved''
cooling with various over-improvement coefficients. In that case the
size distribution will be modified from Fig. \ref{f.siz}: 
it will show in addition
a dominant, divergent contribution of lattice artifacts at small $\rho$.
While this approach has the merit of extracting as much information as
possible, the physically relevant piece is much harder to separate.

\subsection{Further properties of the Instanton Ensemble \label{s.fpi}}

\subsubsection{ Instanton - Anti-Instanton Pairs \label{s.ia}}

Besides instantons and anti-instantons of physical sizes,
the Monte Carlo simulation with Wilson
action produces very many narrow peaks of both positive and negative 
topological charges.
The structure at the level of ``dislocations" disappears very soon in the
cooling ($\sim 5$ sweeps) independently of the lattice spacing, 
while well separated instanton - anti-instanton
pairs of physical size may survive cooling for a while, although eventually 
they have to disappear due to their interaction (pairs are 
not minima of the action).

Because of the unphysical enhancement of short range structure (see \cite{PK} for
an evaluation of this effect) it is difficult to fix precisely the moment
when annihilation of physically relevant I-A pairs starts. Judging, however,
from our data, it  seems that after  20 cooling sweeps the
UV noise at the scale of the cut-off is strongly reduced, independently of $\beta$.
However it is not
easy to design a scaling criterion for the I-A structure since it changes
during cooling and moreover the speed of the changes depends on the
lattice spacing and on the particular cooling algorithm.
Our attitude has been to compare all quantities for
different numbers of sweeps: whenever they are stable with
cooling, as happens for instance with the charge and
size distributions in
Figs.~\ref{f.ch},~\ref{f.siz},  we are confident in considering
them among the
physical attributes of the original Monte Carlo configuration.
Since above 20 sweeps all the changes induced 
by cooling  are mostly due to I-A annihilation, some information about the process
of annihilation can be extracted by looking at the evolution with cooling
of several quantities, like the total number of I's and A's or the 
smallest separation between instanton and anti-instanton. For this argument the stability of instantons under prolonged cooling,
as ensured by the $5Li$ action, is essential.

  The presence of many close I-A pairs in the process of annihilation
imposes certain limits on the reliability of our programs for identifying 
instantons 
(a local structure deformed by the annihilation may show many close peaks 
which should not be identified as separate instantons).
This is taken into account to some extent by the self-duality cut 
but in some cases, mostly for small
numbers of cooling sweeps, this may fail.
Based on the comparison between
different determinations of the instanton size for the $12^4$ and $12^3\times 36$ lattices
(see section~\ref{s.siz}) we  give more than 90$\%$ confidence to our ``peak
program" after 20 cooling sweeps or more.
Since the peak assignment after 5 sweeps has a lower precision we chose not to present
results based on it for such short cooling. 
\footnote{ Note that, even in the continuum,
the problem of identifying all
topological objects is not well-defined: classifying overlapping objects 
of positive and negative topological charges as an I-A pair (to keep) 
or a trivial fluctuation (to smooth out)
becomes arbitrary as their separation becomes comparable to their 
size and the dilute picture breaks down. 
Different degrees of smoothing -- as achieved at
different stages of cooling -- select in some sense different distance 
and overlap scales for the
identification of the topological structure of the configurations. 
In our  ``short cooling" analysis we try to extract as detailed topological 
information as possible
while still preserving a reliable instanton identification.}

What short cooling means from the point of view of I-A annihilation depends 
both on the lattice spacing and on the cooling algorithm.
The annihilation of a pair at a certain physical distance
is slower for smaller $a$ since the effect of cooling a
particular link  extends at most a distance $3\sqrt{2} a$ from it for $5Li$
cooling.  On Table~\ref{t.npeak} we give the number of peaks 
at various cooling stages, compared
with the average action and absolute value of the topological charge. 
A comparison between the $12^4$ ($a=0.12$ fm) and the $24^4$ ($a=0.06$ fm) lattices shows that on the latter about
two or three times  more sweeps are needed to reduce the number of pairs to
the same level as on the former. While dislocations 
annihilate after a number of cooling sweeps roughly independent of $\beta$, physical I-A pairs  annihilate after a number of sweeps which
increases with $\beta$.
Before 20 cooling sweeps for the $12^4$
lattices and 50 sweeps for the $24^4$  lattice the 
abundance of close, annihilating pairs makes the description
of the topological structure uncertain. To compare similar situations   for the two lattices
we present the results based on the peak analysis for 20 (50) sweeps and above.   
Note that the topological charge saturates much earlier (around 5 sweeps, see
also Table~\ref{t.sus}a)
so the discussion above does
not affect at all the determination of the topological susceptibility.

\vskip8mm
\hbox to \hsize{\hfil\vbox{\offinterlineskip
\halign{&\vrule#&\ $#\mystrut$\hfil\ \cr
\noalign{\hrule}
&&& 12^4 \ {\rm  t.b.c.}\  (1a) && {\rm 12^4 \ p.b.c.}\ (1b)
&\cr
&&&\beta=2.4&&  \beta=2.4 
&\cr
&\!{\rm sw}\! 
&& \langle N_{\rm peaks}\rangle\hskip3mm\ \ \ \langle  \hat{S}  \rangle\hskip3mm\ \ \langle  |Q|  \rangle\hskip3mm   \ \ \langle  Q^2  \rangle\hskip3mm   \ \    
&&  \langle  N_{\rm peaks}  \rangle\hskip3mm\ \ \ \langle  \hat{S}  \rangle\hskip3mm\ \ \langle  |Q|  \rangle \hskip3mm  \ \ \langle  Q^2  \rangle\! \ \
&\cr
height4pt&\omit&&\omit&&\omit&\cr
\noalign{\hrule}
height4pt&\omit&&\omit&&\omit&\cr
&20
&&10.6(1)\ \ \  8.42(7)\ \  1.73(7) \ \ 4.5(3)  
&&10.6(2)\ \ \  \ \ 8.3(1)\ \ \ 1.7(1)\ \ 4.5(5)
 &\cr
&50
&&\ \ 5.1(1)\ \ \ 4.48(6)\ \ 1.69(7) \ \ 4.4(3)
&&5.08(8)\ \ \ 4.34(4)\ \ \ 1.6(1)\ \ 4.3(5)
&\cr
&150
&& 2.45(7)\ \ \ 2.20(4)\ \ 1.65(8)\ \ 4.2(3)
&& 2.26(6)\ \ \ 2.15(5)\ \ \ 1.6(1)\ \ 4.0(4)
&\cr
&300
&& 2.02(8)\ \ \ 1.69(7)\ \  1.60(7)\ \ 4.0(3)
&&\ \ 1.7(1)\ \ \ \ 1.6(1)\ \ \ 1.5(1)\ \ 3.9(4)
&\cr
height4pt&\omit&&\omit&&\omit&\cr
\noalign{\hrule}
height4pt&\omit&&\omit&&\omit&\cr
\noalign{\hrule}
&&&12^3\times 36 \ {\rm p.b.c.} \ (1c) &&24^4 \ {\rm t.b.c.}\ (3) &\cr
&&&\beta=2.4&&  \beta=2.6&\cr
&\!{\rm sw}\!
&& \langle N_{\rm peaks}  \rangle\hskip3mm\ \ \ \langle  \hat{S}  \rangle\hskip3mm\ \ \langle  |Q|  \rangle\hskip3mm   \ \ \langle  Q^2  \rangle\hskip3mm   \ \    
&& \langle  N_{\rm peaks} \rangle\hskip3mm\ \ \ \langle  \hat{S}  \rangle\hskip3mm\ \ \langle  |Q|  \rangle \hskip3mm  \ \ \langle  Q^2  \rangle\! \ \
&\cr
height4pt&\omit&&\omit&&\omit&\cr
\noalign{\hrule}
height4pt&\omit&&\omit&&\omit&\cr
&20
&&31.7(4)\ \ \  25.3(3)\ \ \ 2.9(2)\ \ 13(1)
&&63.1(5)\ \ \  30.7(7)\ \ \ 1.7(2)\ \ 4.4(8)
&\cr
&50
&&15.4(3)\ \ \ 13.2(2)\ \ \ 2.8(2)\ \ 12(1)
&&  15.7(5)\ \ \ 12.8(6)\ \ \ 1.5(1)\ \ 4.3(7)
&\cr
&150
&&\ \ 6.6(3)\ \ \ \ 5.9(2)\ \ \ 2.7(2)\ \ 12(1)
&&\ \ 6.7(3)\ \ \ \ 6.1(3)\ \ \ 1.6(2)\ \ 4.2(7)
&\cr
&300
&&\ \ 4.5(2)\ \ \ \ 4.0(2)\ \ \ 2.6(2)\ \ 11(1)
&&\ \ 3.9(2)\ \ \ \ 3.8(2)\ \ \ 1.6(2)\ \ 4.0(6)
&\cr
\noalign{\hrule}}}\hfil}
\vskip3mm
{{\noindent Table \ref{t.npeak}: Cooling behavior of the average number of peaks
$\langle N_{\rm peaks}\rangle$,
$\langle \hat{S}\rangle$, $\langle |Q|\rangle$ and $\langle Q^2\rangle$. 
The (1c) data for $\langle N_{\rm peaks}\rangle$,                         
$\langle \hat{S}\rangle$, $\langle |Q|\rangle$ should be rescaled by a factor $\sqrt{N_s/N_t}$   
and for $\langle Q^2\rangle$ by a factor $N_s/N_t$
in order to be compared with the others.
 }\par}
\label{t.npeak}
\vskip5mm

In asking which are the typical features (size, location) of the physically relevant pair structure we can, to
a certain extent, look for an answer by comparing
the distributions, say, after 20(50) and after 300 cooling sweeps.
Although between the two cooling stages about $80\%$ of the (anti-)instantons
have annihilated (see
Table~\ref{t.npeak})  the
shape of the size distributions is very similar. This is clearly visible on the normalized distributions of Fig.~\ref{f.siz} and suggests that the size distribution
given there holds also for the
instantons and anti-instantons appearing in pairs, which have not
annihilated in the very early stages of cooling. Since this
distribution scales correctly with the cut off this is simultaneously a
consistency check that no unphysical structure was present at 20(50) sweeps.
This bound is surely an overestimate, as has been argued above. We have, however, no
reliable way of characterizing the physical structure bound in close
pairs which annihilate between 5 and 20(50) sweeps.

\subsubsection{ Distance Distributions \label{s.dis}}

In Fig.~\ref{f.dis} we present distance distributions for alike and opposite
charge objects. The solid line is the 4-dimensional volume factor, that is it
 corresponds to a homogeneous distribution of instantons over the lattice
(the shape is due to the finite size of the hyper-cubic, periodic lattice). The figure shows little
difference with a homogeneous distribution. After 300 sweeps most of the I-A pairs
have disappeared and are thus not included in the figure.

In Fig.~\ref{f.dism} we present distributions of distances to the closest alike and opposite charge objects. The center of the distribution shifts toward
larger distances with longer cooling as a result of pair annihilation. 
For the $24^4$ lattice we observe also a small peak in the alike distribution
at short distances ($\simeq 3a$) and short cooling, which disappears with
further cooling. A much stronger peak at small distances (about 1 to $2a$)
has been observed by \cite{MS}. In our opinion structure at the scale of
the cut-off is unphysical and in fact  the cuts we impose in the peak assignment
(see section \ref{s.mc}) take care of most of it. The remnant at $3a$ may have
physical significance. We believe however that it is to be attributed to  a
bad assignment of peaks.  A wide, deformed instanton could also
trigger such an effect, but it should not depend then on the amount of cooling.
We  think that the annihilation of I-A pairs is the dominant effect:
in the course of pair annihilation, transient deformed structures can 
appear which lead to multiple alike-charge peaks. 
These then automatically produce a temporary
increase in the small distance part of the alike distribution,
an effect which is removed by
further cooling. The cuts we imposed from self-duality 
take care of most of this effect but still part of it has remained.

An interesting information provided by 
Table~\ref{t.npeak} is that the number of pairs, $N_{\rm pairs}$, observed after short 
cooling -- 20 sweeps for the $12^4$ lattice and 50 sweeps for the $24^4$
lattice -- is significantly larger than the value corresponding to
a binomial distribution with equally probable, uncorrelated I and A events, which would give
\be
2\langle N_{\rm pairs} \rangle  + \langle |Q| \rangle 
\equiv \langle N_{\rm peaks} \rangle = \langle Q^2 \rangle .   
\label{e.pois}
\ee
 Simultaneously we observe
$ \langle \hat{S} \rangle / \langle N_{\rm peaks} \rangle  < 1$ (see also Fig.~\ref{f.npeak}
in next section). 
The distributions at 20(50) sweeps seem in fact compatible with binomial laws for
uncorrelated I, A  (with equal probabilities such that $\langle N_{I}+N_{A}\rangle=\langle Q^2 \rangle$, see eq. (\ref{e.pois})) {\em and} I-A events
(accounting for the surplus of I-A pairs, $\langle\Delta N_{\rm pairs}\rangle = (\langle N_{\rm peaks}\rangle-\langle Q^2\rangle)/2$), see the histograms
in Fig.~\ref{f.dism} \footnote{
For comparison with the actual MC data, we generated spatial configurations
of charges with the same susceptibility $\langle Q^2\rangle$ 
and number of peaks $\langle N_{\rm peaks}\rangle$ as
follows. Pairs of charges are placed on the lattice with an average density
$d_p$, at locations $x$ and $y$ chosen independently with uniform probability. They
do not contribute to the susceptibility, but give rise to $2 d_p V$ peaks on
average. In addition, single charges are added with average density $d_s$, at
random locations, contributing an additional $d_s V$ to the number of peaks,
so that $\langle N_{\rm peaks} \rangle = (d_s + 2 d_p) V$.  If the sign of these single charges is
chosen randomly, the susceptibility will be $\langle Q^2\rangle = d_s V$. This is
appropriate if $\langle N_{\rm peaks}\rangle > \langle Q^2\rangle$, ie. if there is a surplus of pairs over a
binomial distribution. If instead there is a defect of pairs, as for long
cooling, the sign of the single charges is chosen ``en bloc'' for each
configuration. The susceptibility is then $\langle Q^2\rangle  = d_s V + (d_s V)^2$. In
both cases, the location of each charge is {\em uncorrelated} with that of
the others.}.
 Due to the annihilation of pairs during cooling this distribution 
is modified and  after long cooling evolves towards   
a binomial distribution for only one type of charges
(as after 300 sweeps for lattices (1a,b)).
These facts taken together 
are compatible with an I-A interaction, which would be responsible both  
 for enhanced  pair creation in the Monte Carlo 
production
 and for
pair annihilation during cooling. This interaction is surely dependent on 
the overlap between opposite charge objects.  
We can characterize the overlap for I-I or I-A pairs as 
\be
{\cal O} = \frac{\rho_1+\rho_2}{r_{12}}
\label{e.ovl}
\ee
where $\rho_1$ and $\rho_2$ are the sizes of the objects in the pair and
$r_{12}$ their mutual distance.
We obtain for the average over (1a,b) lattices
after 20 sweeps
$\langle {\cal O}\rangle \sim 1.7$ for I-A pairs and
$\langle {\cal O}\rangle \sim 1.8$ for I-I pairs.
 On the (3) lattice after 50 sweeps $\langle {\cal O}\rangle \sim 1.7$ for I-A pairs and
$\langle {\cal O}\rangle \sim 2.0$ for I-I pairs.
The overlap we observe is rather large, suggestive of a percolating
polymer structure rather than a dilute gas.

Since the action is lowered by the interaction,
which increases with the overlap,
we would expect increased production of opposite pairs with
larger overlap.
Surprisingly however, there is no indication after 20(50) cooling
sweeps of a larger overlap for opposite charge than for alike charge pairs,
which we would naively expect to result from enhanced production of close
I-A pairs in the original Monte Carlo data.
Hence we can only say that the I-A interaction favors
configurations with a surplus of I-A pairs over a naive
binomial distribution, but with a rather
homogeneous, uncorrelated distribution of I's and A's over the lattice.
Since cooling also preferentially
destroys pairs with large overlap we cannot exclude, however, that close
I-A pairs were more abundant in the uncooled configurations.

\subsubsection{Fractional Charge Instantons \label{s.fci}}

It has been recently pointed out \cite{GAM} that instantons with topological charge $1/2$
might play a relevant role in the SU(2) vacuum. The presence of
such structure could modify the general picture of the instanton
ensemble. In particular it will change the statistical relation between
susceptibility and number of peaks, eq. (\ref{e.pois}), since 
such objects carry only half the topological charge and action 
of ordinary instantons.
Part of the analysis in \cite{GAM} to detect them
was based on a determination of the number of peaks in the energy density compared
with the average value of the action. If instantons of charge 1 are the dominant structures
in the cooled Monte Carlo configurations one expects that the average number of peaks
will coincide with $\langle\hat{S}\rangle$. Instead in Ref.~\cite{GAM} an excess in the number of peaks was found.
We have pursued a similar analysis over our configurations. In Table~\ref{t.npeak}
we have given the number of peaks compared with $\langle\hat{S}\rangle$  as a function of cooling sweeps.
There is an excess of $N_{\rm peaks}$ over $\langle\hat{S}\rangle$ but what we observe 
amounts in general to $20\%$ or less, at short cooling, and decreases fast with 
cooling, in such a way that for 300 sweeps both numbers are
roughly compatible within errors. This indicates that if such structure was
there initially it should have appeared mostly in opposite charge
pairs which have annihilated after long cooling.  
It is particularly difficult to distinguish a $Q = \pm 1/2$ pair from an
ordinary I-A pair and the above effects could be due also to
the latter.
The presence of several topological objects in our physical box
prevents the integration of the topological charge density over all space:
when these objects are large compared to their separation and form a 
dense gas, the total charge carried by each object becomes ambiguous.
After 300 sweeps we looked directly into some of the few
configurations for which there was an excess in the number of peaks.
We found some suggestion of the presence of $Q=1/2$  objects,
but we do not know a good way of identifying and separating them further
from some other kind of structure like wide instantons, a problem already
mentioned in \cite{GAM}.  
In any case, the effects under discussion
are rather close to the uncertainty level of $\sim 10\%$  we assign
to our programs for identifying peaks.

 In Fig.~\ref{f.npeak} we present the distribution
of $\mu_\rpp= \hat{S}/ N_{\rm peaks}$.
If the $Q=1/2$ were the
dominant objects the points would concentrate mostly around
$\mu_\rpp=1/2$ while for ordinary instantons they will concentrate
around $\mu_\rpp=1$.  Our results show a clear tendency  
towards $\mu_\rpp=1$.

On the basis of these results we conclude no clear indication for
the presence of such objects in our MC configurations.     
Note, however, that due to the periodicity of our boundary 
conditions in spatial directions,
fractional charge instantons cannot appear alone.
It is possible that to see a clear effect of these objects   
under these conditions larger volumes than the ones used
in our simulations are needed (the ones employed in \cite{GAM}
are almost twice as large as ours). This is an interesting proposal
which deserves further investigation.

\subsection{Density - Density Correlations}

Using the improved charge operator and action allows to
measure meaningful density - density correlations even after a small number
of cooling sweeps. The extraction of the instanton features from
these data is more involved and therefore matter for a separate
investigation. We show here part of these data on Figs. \ref{f.qq2} and
\ref{f.qq5} only to
indicate the compatibility with results of the previous analysis. 
We note by $q_+(x)$ and $q_-(x)$ ${\rm max}(q(x),0)$ and ${\rm min}(q(x),0)$ respectively.

Several features are noticeable:\par
\begin{itemize}
\item 
For short cooling $\langle q(0) q(r) \rangle$ goes smoothly to zero as $r$ increases, 
as expected, indicating a rather homogeneous distribution of instantons
and anti-instantons. For long cooling the tail represents a more or
less homogeneous distribution of a few instantons {\it or} anti-instantons, 
since practically all pairs have meanwhile disappeared.
\item
 As $r \rightarrow \infty$, 
${\rm lim} \langle q(0)_+ q(r)_+\rangle = {\rm lim} \langle q(0)_- q(r)_-\rangle$. This is a consequence
of the evenness of the $q$-distribution.
For short cooling and as a consequence of the homogeneous distribution
of instantons and anti-instantons we see in addition
${\rm lim} \langle q(0)_+ q(r)_+\rangle= - {\rm lim} \langle q(0)_+ q(r)_-\rangle 
= \frac{1}{4} {\rm lim} \langle |q(0)| |q(r)|\rangle$, within statistical errors. 
For long cooling, however, as a consequence of the disappearance of most I-A pairs 
${\rm lim} \langle q(0)_+ q(r)_-\rangle\approx 0$, and ${\rm lim} \langle q(0)_+ q(r)_+\rangle=\frac{1}{2} {\rm lim} \langle |q(0)| |q(r)|\rangle$.
\item 
The peak of the $q-q$ correlations near the origin is caused by
the typical size of the instanton. It can be used 
to extract information about the instanton size \cite{MIT1}. 
For comparison we present (dotted line in Fig. \ref{f.qq5})
the distribution obtained with one instanton obeying the continuum ansatz eq. 
(\ref{e.Qc}), with 
$\tilde{\rho}_\mu = \tilde{\rho}_M \equiv 0.43$ fm, rescaled to match
the measured value of the $\langle q(0)q(r)\rangle$ correlation at the origin.
The moderate agreement shows the limitations of a one-instanton ansatz
in interpreting the density correlations (see below for a more
realistic ansatz).
\item 
Comparing the $q-q$ and $|q|-|q|$ correlations we see clearly the
minimal distance between opposite charges, in agreement with vanishing
of the $q_+-q_-$ correlation below this distance. 
\end{itemize}

To help us interpret our Monte Carlo data on density-density correlations,
we have constructed some ``synthetic" data from the simplest ansatz:
we place on the lattice instantons obeying the continuum ansatz  
(with periodicity effects accounted for by adding the 80 nearest
replicas). 
These ``synthetic" instantons do not interact.
They are distributed homogeneously (with no spatial correlations)
as described in section \ref{s.dis}, with densities of instantons and of
I-A pairs adjusted to match
the susceptibility and the number of peaks measured in the Monte-Carlo. 
The size distribution for these continuum instantons is chosen according 
to the measured size distribution. 
We can then compare our density-density correlations with those produced
by this parameter-free ansatz. The ansatz results are given by the solid
curves in Figs.  \ref{f.qq2}, \ref{f.qq5}.
The agreement with the actual Monte Carlo data is excellent (it becomes
almost perfect if one allows for an overall normalization factor, to
account for, e.g., the small mismatch of $\langle Q^2\rangle$ and 
$\langle N_{\rm peaks}\rangle$). This
validates
a posteriori several aspects of our approach: deviations of our lattice
instanton fluid from a linear superposition of continuum ansatz are very
small; instanton sizes have been determined reliably. 
Only the correlations of the action density at short cooling
are not so well reproduced by our non-interacting ansatz. 
This could be expected, since our ansatz
assigns an action of $8 \pi^2$ per charge, whereas the action we
actually measure at 20 cooling sweeps is smaller (see Table 3.4.1). 
This is our
first signal for an instanton-anti-instanton interaction. Its influence
on the spatial charge distribution, however, is too weak for us to detect
directly here.

The features of the density-density correlations listed above are preserved 
when going to smaller  
scales, as indicated by comparing Figs. \ref{f.qq2} and \ref{f.qq24} 
(notice that the normalizations are different, since the number of peaks is larger 
on the $24^4$ lattice than on the $12^4$ lattice, see Table \ref{t.npeak}).
It would be very interesting to compare these correlations with the 
predictions of phenomenological models, and to observe the effect of
temperature and of dynamical fermions. We are starting work in these directions.

The correlation $\langle q(0) q(r) \rangle$ of the topological charge density cannot
be directly compared with the density-density correlation of a classical
fluid made of positively and negatively charged particles.
We do not deal with point-like or hard-core particles and the $q-q$
correlation smears the information about the center of the instanton.
We could apply a de-convolution, to reconstruct the correlation
for an instanton to be at $r$, given the center of another one at the
origin. 
Instead we measure such correlations directly, by placing a charge $Q = \pm 1$ 
at the location of the peak of $q$. The correlations $\langle \hat{q}(0) \hat{q}(r) \rangle$
thus obtained after this ``lumping'' of the topological charge are 
displayed in Fig. \ref{f.qqd}. Since there are few peaks at small
distances (in fact we identify peaks at distances up to 1 in each direction),
there is no contribution to these correlations here. 
By comparing these data with those for the full densities we
clearly see the effect of the size of the instanton at short distances.
The $\hat{q} -\hat{q}$ correlation is zero both at small and at large distances, 
corroborating the smooth decrease $\langle q(0) q(r) \rangle$ with $r$,
and confirming the overall absence of structure in our instanton fluid. 
The small difference between the $\hat{q}_+ -\hat{q}_-$ and the $\hat{q}_+ -\hat{q}_+$
in the intermediate region may be a signal of pair annihilation. This 
would explain also the slight
increase of $|\hat{q}|-|\hat{q}|$ in the same region. For further conclusions
the data are too noisy yet. 

As a final check for the appearance of regular (``crystalline") structures, we
analyzed the Fourier Transform of the density-density correlations. The
spectrum does not show the typical peaks produced by a crystal.
In agreement with our observations from the distance distribution,
we conclude that there is a strong indication for a homogeneous,
spatially uncorrelated distribution of instantons over the lattice.

\section{ Conclusions \label{s.conc}}

The goal of our study is to identify classical objects (instantons) on the
lattice. For this purpose, it is appropriate to use a classically improved
lattice action and topological charge operator. Our 5-loop operator has a
tree level discretization error starting with ${\cal O}(a^6)$, which is 
fine-tuned to minimize violations of scale-invariance. We use this 
operator both for measurements, and for cooling, which rapidly isolates 
the topological components of the QCD vacuum.
With this improved cooling algorithm we have thus a straightforward
 method to remove UV noise and dislocations
without modifying the topological
structure at physical scales (in so far as it is not affected
by finite size effects or pair annihilation). The excellent scale invariance 
properties of our cooling action allows us to identify unambiguously
the physical structure 
of the original configuration which carries the net topological charge,
since instantons are completely stable under any amount of cooling.
A scaling analysis of the results at various levels of cooling
shows that already before 20 improved cooling sweeps UV noise and
dislocations have disappeared and only physical structure remains,
including I-A pairs.
The improved cooling, being a minimization procedure,
does not preserve I-A pairs, but since there are no other modifications
of physical instantons during cooling we can estimate the contribution
of the pairs by comparing results at different cooling stages.
The improved charge operator gives smooth density
data already on still rough configurations, data which can be used
also for correlation analysis. The total charge comes out
an integer within ${\cal O} (1\%)$ after 5 - 10 sweeps and remains stable thereafter.

The dislocation cut off is fixed in lattice units ($\tilde {\rho}_0 \simeq
2.3a$) and therefore
will not affect the physical scales for $a$ small enough (such that
$\tilde {\rho}_0$ falls below these scales). The finite size effects
begin to affect instantons of sizes above $L/3$. Therefore
 we should take care to use lattices
larger than three times the typical instanton size.

The identification and description of the structure in the charge and
energy densities
can successfully be based on the continuum
ansatz for instantons, especially if one takes into account
anisotropy effects induced by quantum fluctuations and clustering effects. Uncertainties in
the identification procedure affect either
large instantons deformed by finite size effects or unstable structures like
I-A pairs in late stages of annihilation.

Using improved cooling we have analyzed the topological content
of the $SU(2)$ theory, especially the topological susceptibility and the
instanton size and distance distributions. Since the decisive criterion for the physical
significance of the results is the correct scaling with the cut off in approaching
the continuum limit, we worked at different scales. We checked that under a rescaling of $a$ by a factor of two the susceptibility
and the relevant part of the size distribution both scale very well. We also verified that
the lattices we used (size 1.4 fm, $12^4$ and $24^4$) were large
enough not to introduce finite size effects for the relevant physical sizes. Our results for the susceptibility and  the size distribution 
are stable under cooling, which only affects the distance distribution 
as the result of pair annihilation.

The value of the susceptibility, 200(15)MeV, agrees well with the Witten-Veneziano formula.
The most likely instanton size turns out to be $0.43(5)$ fm, with a full width of about $0.47$ fm.
This size is larger than the one proposed for the instanton
liquid models $\rho \sim 1/3$fm \cite{models}. The distance
distribution is similar to that of randomly placed objects, and
there is no indication for a crystalline structure.
The structure of configurations at scales above the threshold
$\tilde{\rho}_0 \simeq 2.3a$ - which we argued to be the physically relevant
ones in what concern topological excitations - appears to be
a random mixture of instantons and anti-instantons of sizes compatible with an
effective semi-classical distribution and with a homogeneous spatial
distribution. With short cooling (20 (50) sweeps for the $12^4$
($24^4$) lattices), however, we observe more pairs than would be compatible with 
uncorrelated binomial distributions for instantons and anti-instantons,
suggesting that due to pair interaction the production of opposite charge
pairs is enhanced. No significant correlation between the
locations of A and I objects is seen here,
but we cannot exclude the presence in the original configurations
of closely bound I-A pairs which annihilated early during the cooling.
The relative overlap  eq.~(\ref{e.ovl}) of about 1.5 - 2 among closest
objects makes the dilute description questionable at the scale of short cooling -- 20 sweeps on the (1a,b) lattices.
The density of instantons there is about
$2-3/{\rm fm}^4$ and the fraction of volume they occupy, counting ${{\pi^2} \over 2}\rho_M^4$
for each, is about $30-50\%$. The long cooling
regime is characterized by the absence of pairs and by a binomial distribution
for the surviving type of charges, with a density set by the 
scaling susceptibility.  Note that these numbers depend on the criterion
for identifying instantons in the presence of I-A pairs, which is subject 
to an ambiguity existing already in the continuum.
Finally, there is no clear indication for fractionally charged instantons, but for this analysis larger volumes might be necessary.

{\bf Acknowledgments}: FOM and DFG support for MGP and DFG support for IOS is
thankfully acknowledged, likewise provision of computer facilities by the
Universities of Heidelberg and Karlsruhe. It is a pleasure to thank Pierre van Baal for 
discussions and suggestions and Rajamani Narayanan for providing us with
their results about the topological 
susceptibility. MGP would like to thank Tony Gonz\'alez-Arroyo
for many fruitful discussions about fractional charge instantons.

\newpage

\begin{figure}[htb]
\vskip2cm
\vspace{7.cm}
\includegraphics{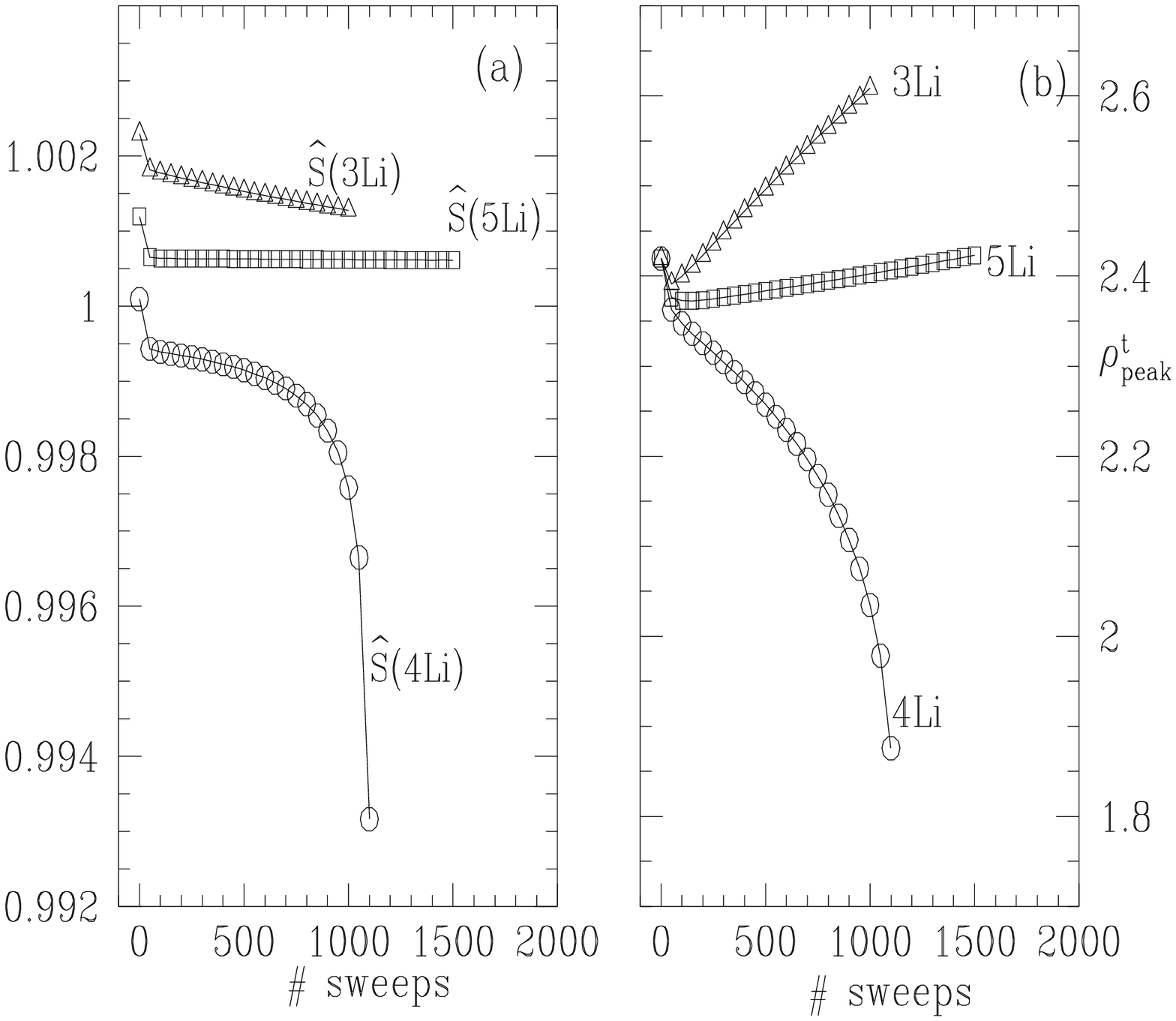}
\caption{ Evolution of: (a) action, $\hat{S}$,  and (b) size, $\rho_{\rp}^t$ (see eq. (\ref{e.rtpk})),
 with the number of cooling sweeps for various choices of 
the improvement coefficient $c_5$ (i.e. $c_5=1/10$, $c_5=0$ and $c_5=1/20$ for $3Li$, $4Li$
and $5Li$ respectively).
\label{f.iact}}
\end{figure}

\vskip3cm
\begin{figure}[htb]
\vspace{4.cm}
\includegraphics{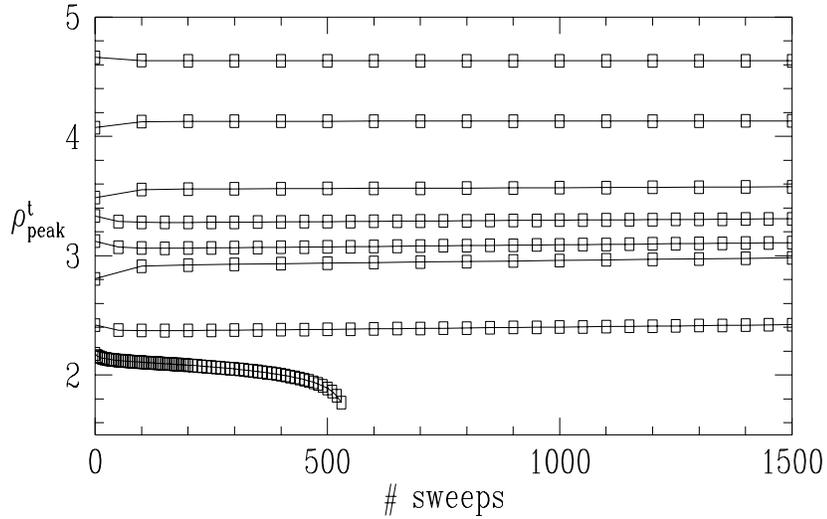}
\caption{Evolution of instanton size under $5Li$ improved cooling for  various
initial configurations. Notice the stability of the size with cooling for instantons above the
threshold $\tilde{\rho}_0=2.3a$.
\label{f.icoo}}
\end{figure}
\newpage

\begin{figure}[htb]
\vskip2cm
\vspace{6.4cm}
\includegraphics{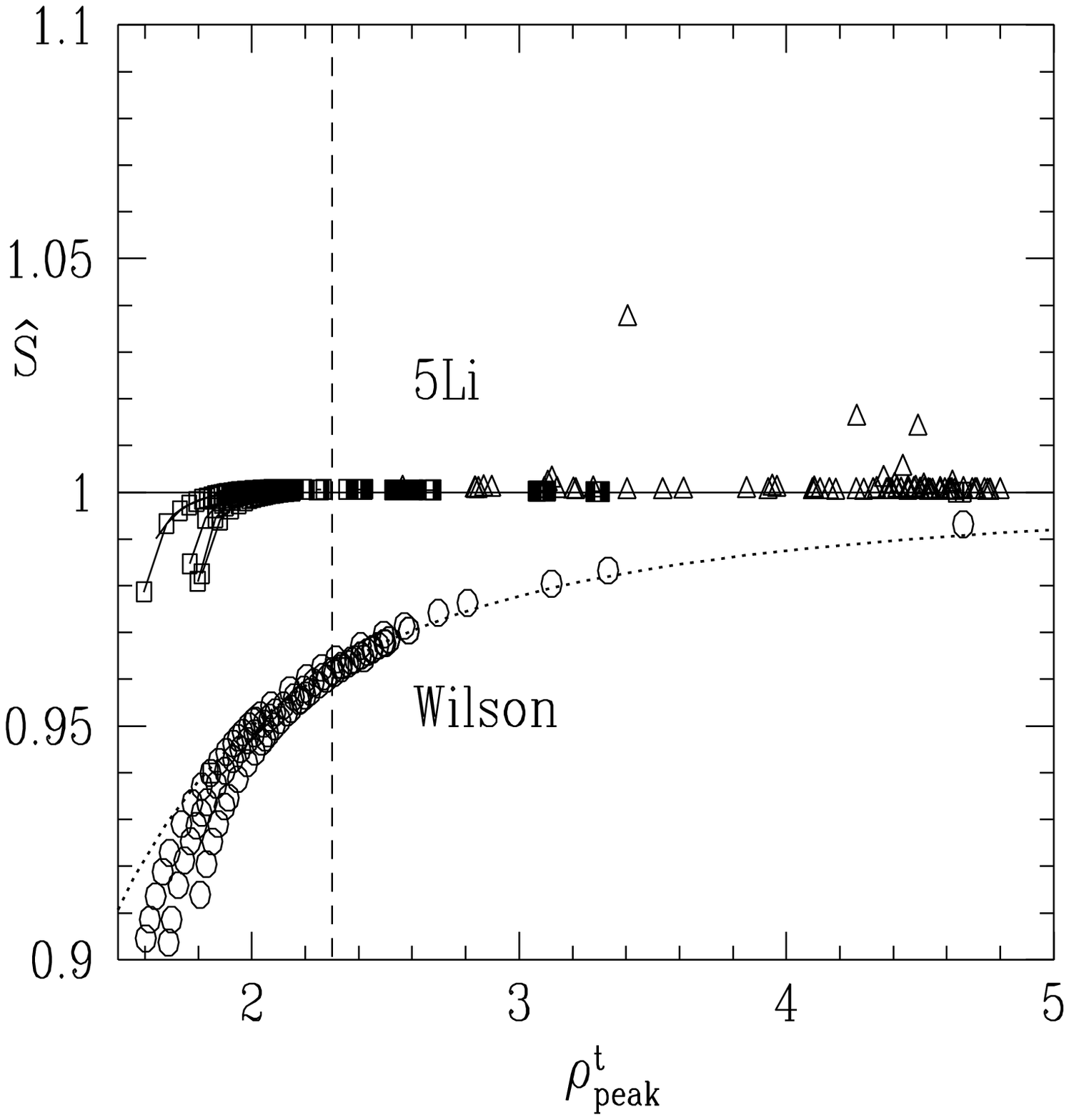}
\includegraphics{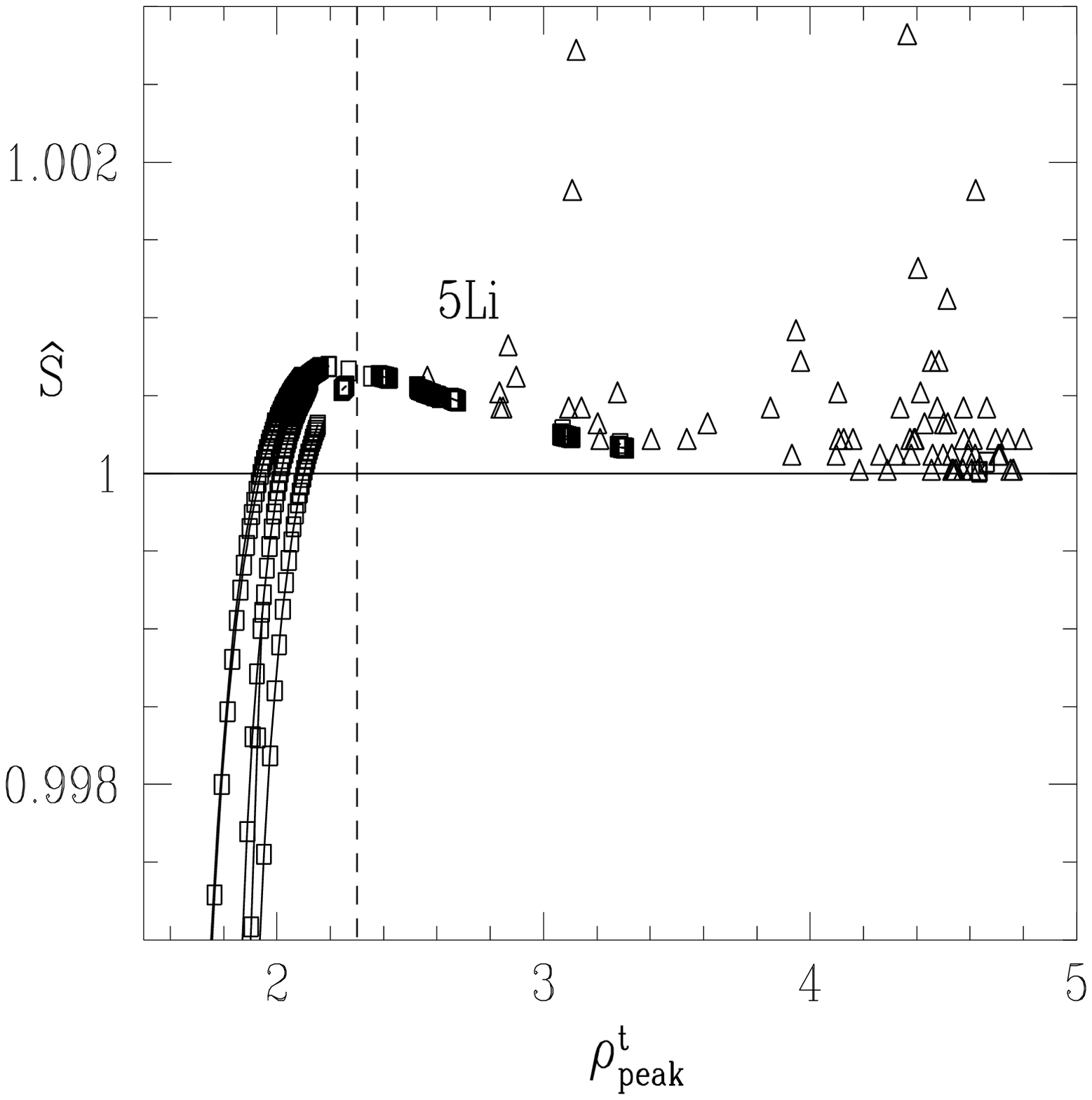}
\caption{ Dependence of the $5Li$ action (squares) and Wilson action (circles)
on the instanton size.  The triangles are data coming from Monte Carlo configurations  generated
with the Wilson action and cooled with 300 sweeps of $5Li$ cooling.  The vertical dashed line indicates the location of
the threshold $\rho_0=2.3$. The dotted curve is the value of the Wilson action for
an instanton of size $\rho$ calculated at tree level to ${\cal O}(a^2)$; it shows
the deviation from scale invariance of the Wilson action.
\label{f.ienv}}
\end{figure}
\vskip4cm

\begin{figure}[htb]
\vspace{3cm}
\includegraphics{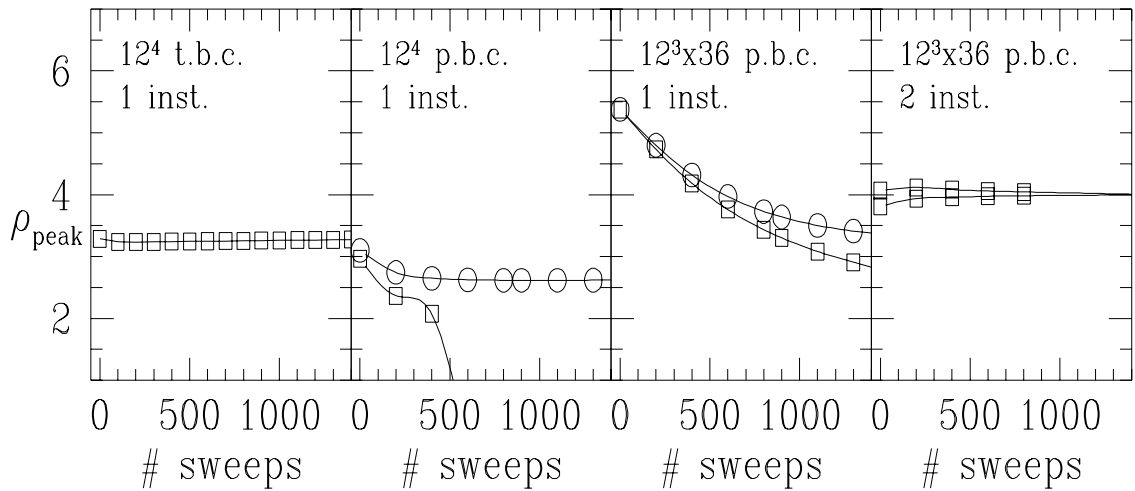}
\caption{Behavior under cooling of configurations with one instanton.
We show the dependence on lattice geometry and boundary conditions.
Squares and circles correspond respectively to $5Li$ improved cooling and
over-improved cooling with $\epsilon=-1$. We also present for 
comparison a 2-instanton configuration.
}
\label{f.q1}
\end{figure}

\newpage

\vskip2cm
\begin{figure}[htb]
\vskip2cm
\vspace{15cm}
\includegraphics{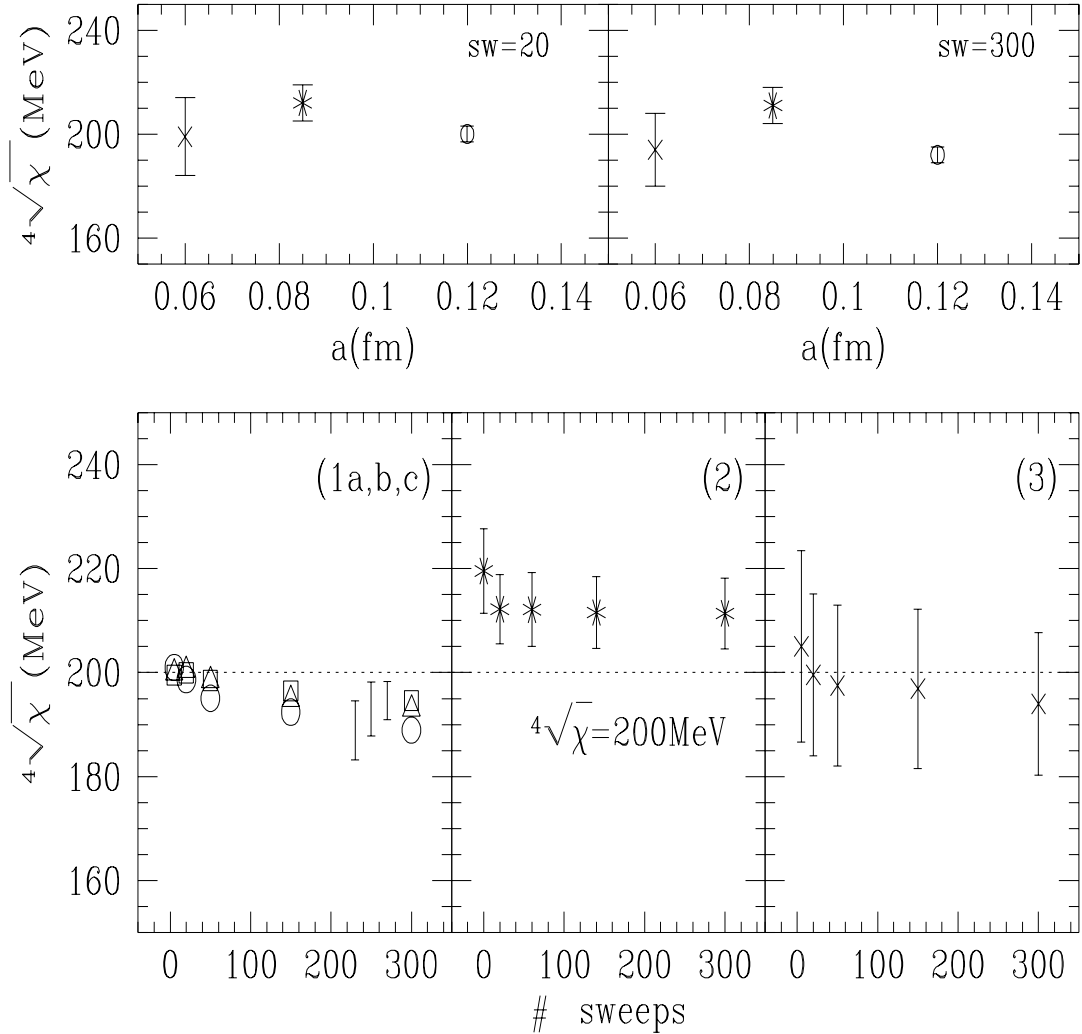}
\caption{Values of the topological susceptibility for the various lattices:
 t.b.c. (1a)(squares), (2)(stars) and (3) (crosses),
and  p.b.c. (1b) (triangles) and (1c) (circles) -- see Table \ref{t.dat}.
In lattices (1a,b,c) we have suppressed the error bars to make the figure
clearer, the bars are an indication of the magnitude of the error for,
from right to left, (1a,b,c).
In the upper plot we present the dependence of the topological
susceptibility with lattice spacing after 20 and 300 cooling sweeps.
The point at $a=0.12$fm is an average over lattices (1a,b,c).
}
\label{f.sus}
\end{figure}
\newpage
\vskip3cm
\begin{figure}[htb]
\vspace{12cm}
\includegraphics{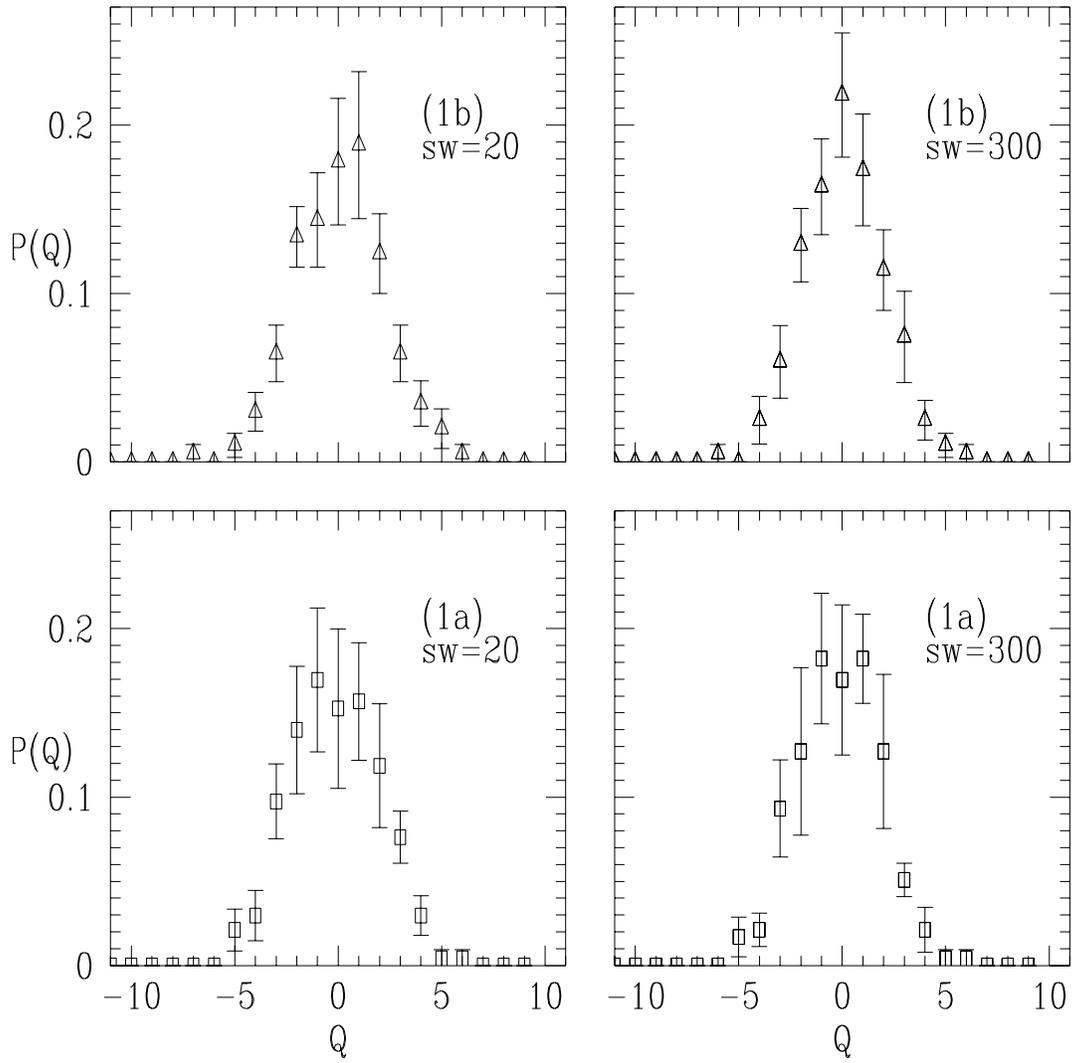}
\caption{Charge distribution after 20 and 300 $5Li$ cooling sweeps for the lattices:
(1a) (squares) and (1b) (triangles) -- see Table \ref{t.dat}.}
\label{f.ch}
\end{figure}

\newpage

\vskip6cm
\begin{figure}[htb]
\vspace{16cm}
\includegraphics{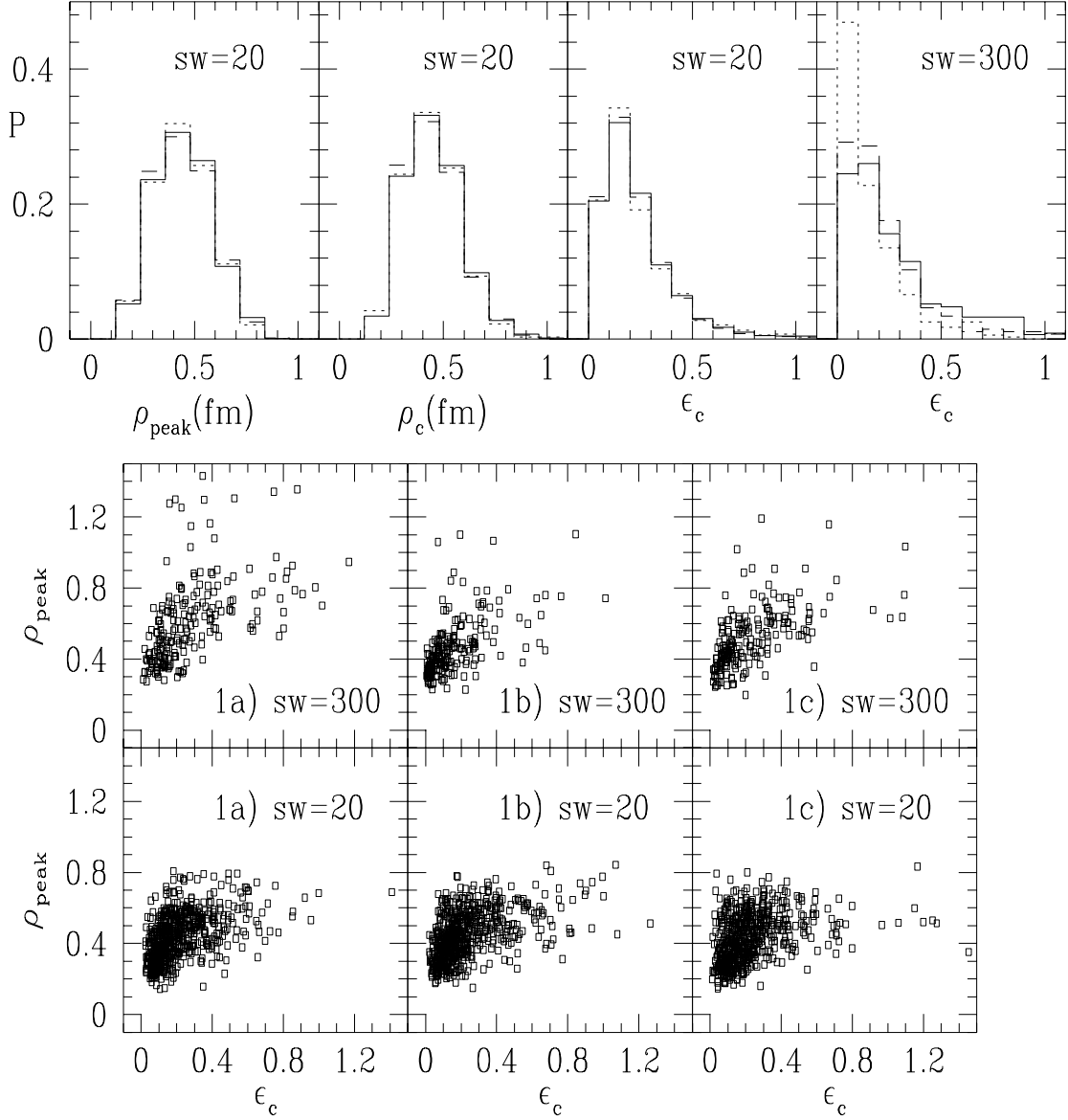}
\vskip2cm
\caption{ Upper plot: size distributions after 20  cooling sweeps
extracted from peak size $\tilde{\rho}_{\rp}$, eq. (\ref{e.rpk}), and 
``center and nearest-neighbor" fit $\tilde{\rho}_c$, eq. (\ref{e.rcpk}).
We also show the histogram of excentricity $\epsilon_c$, eq. (\ref{e.ef}) and (\ref{e.rcpk}), after 20 and 300 cooling sweeps.
The continuous, dotted and dashed histograms correspond to lattices (1a), (1b) and (1c) respectively
(see Table  \ref{t.dat}).
 Lower plot: dependence of the excentricity $\epsilon_c$ on instanton size,
lattice geometry and boundary conditions.}
\label{f.rfp}
\end{figure}
\newpage

\vskip8cm
\begin{figure}[htb]
\vspace{17cm}
\includegraphics{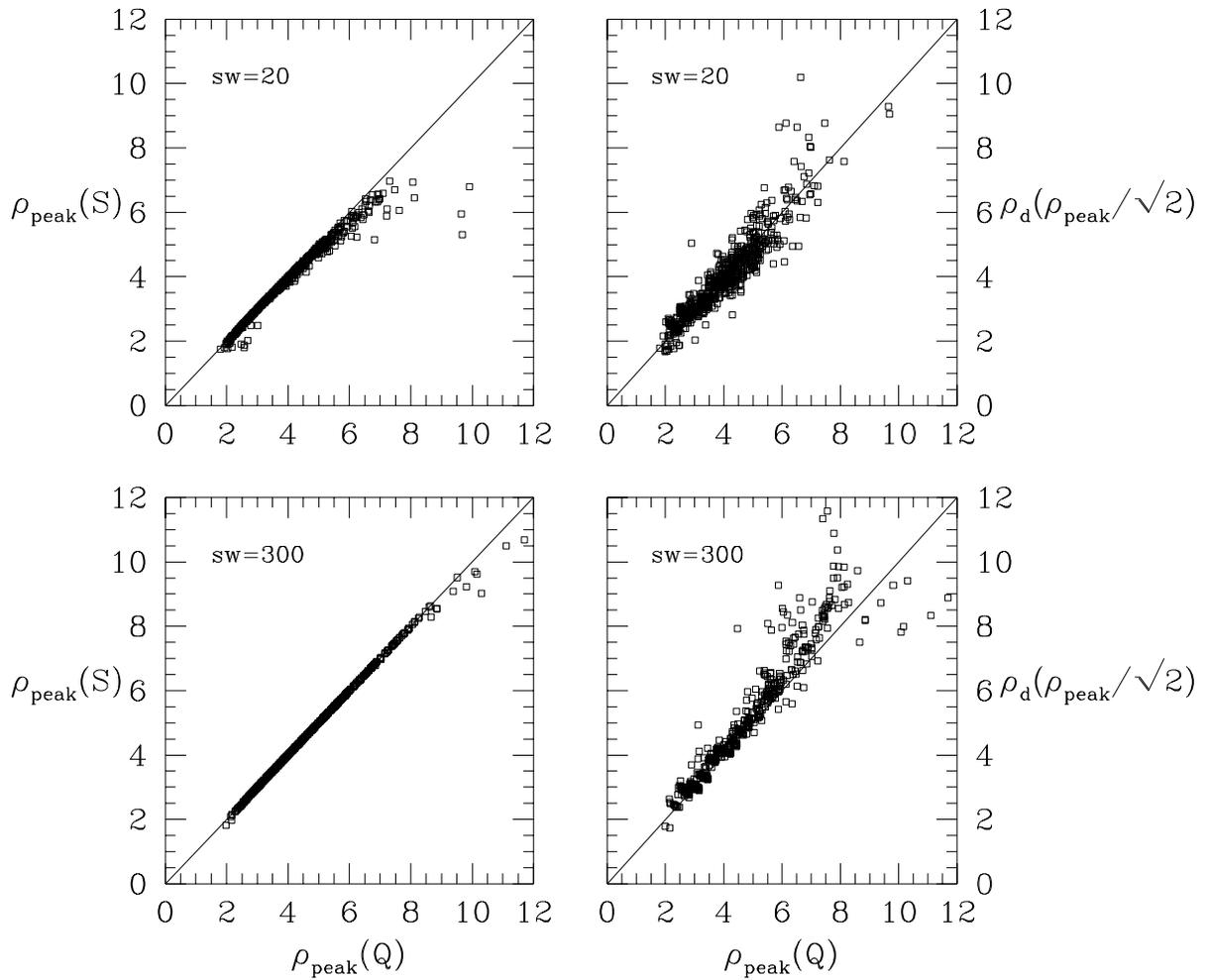}
\caption{Comparison between different size definitions for:
upper plot: 20 cooling sweeps; lower plot: 300 cooling sweeps.
Data are taken on (1a,b,c) lattices.}\label{f.rqs}
\end{figure}
\newpage

\vskip2cm
\begin{figure}[htb]
\vspace{5cm}
\includegraphics{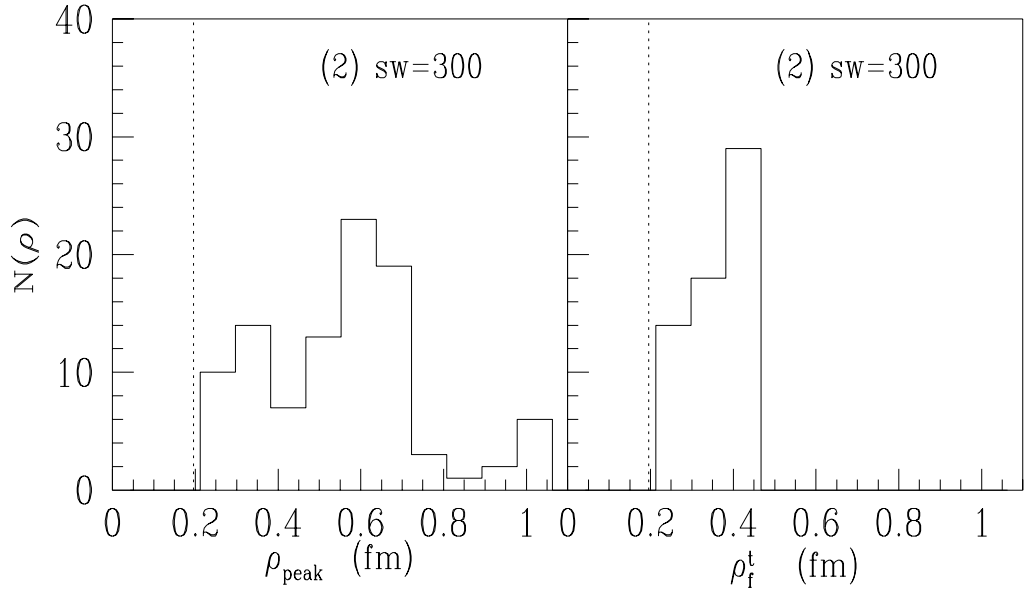}
\caption{Comparison between instanton size distributions
extracted from $\tilde{\rho}_{\rp}$ and $\tilde{\rho}_{\rm f}^t$. 
Data are obtained by cooling for 300
sweeps configurations obtained from a 
Monte Carlo simulation on  a $12^4$ lattice, with $\beta=2.5$ and twisted boundary conditions.
$N(\tilde{\rho})$ denotes the number of instantons with size $\tilde{\rho}$.
}
\label{f.b25}
\end{figure}
\vskip3cm
\begin{figure}[htb]
\vspace{8cm}
\includegraphics{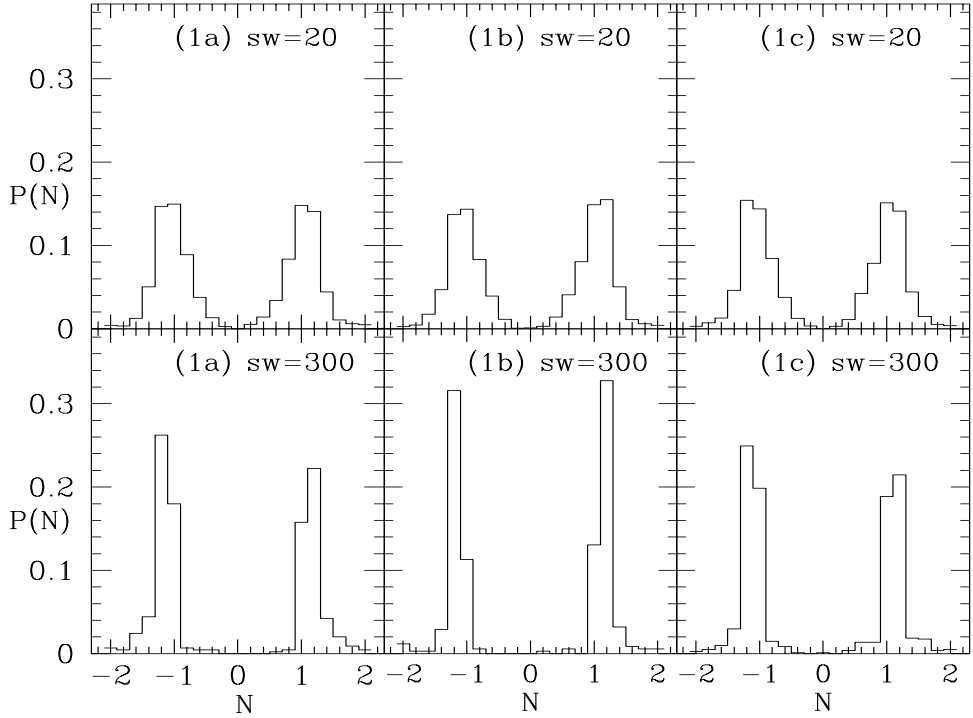}                 
\caption{Histogram of normalizations (see eq. (\ref{e.Qc})).                     
Upper plot: 20 cooling sweeps; lower plot: 300 cooling sweeps.}\label{f.norm}
\end{figure}

\newpage

\begin{figure}[htb]
\vspace{17cm}
\includegraphics{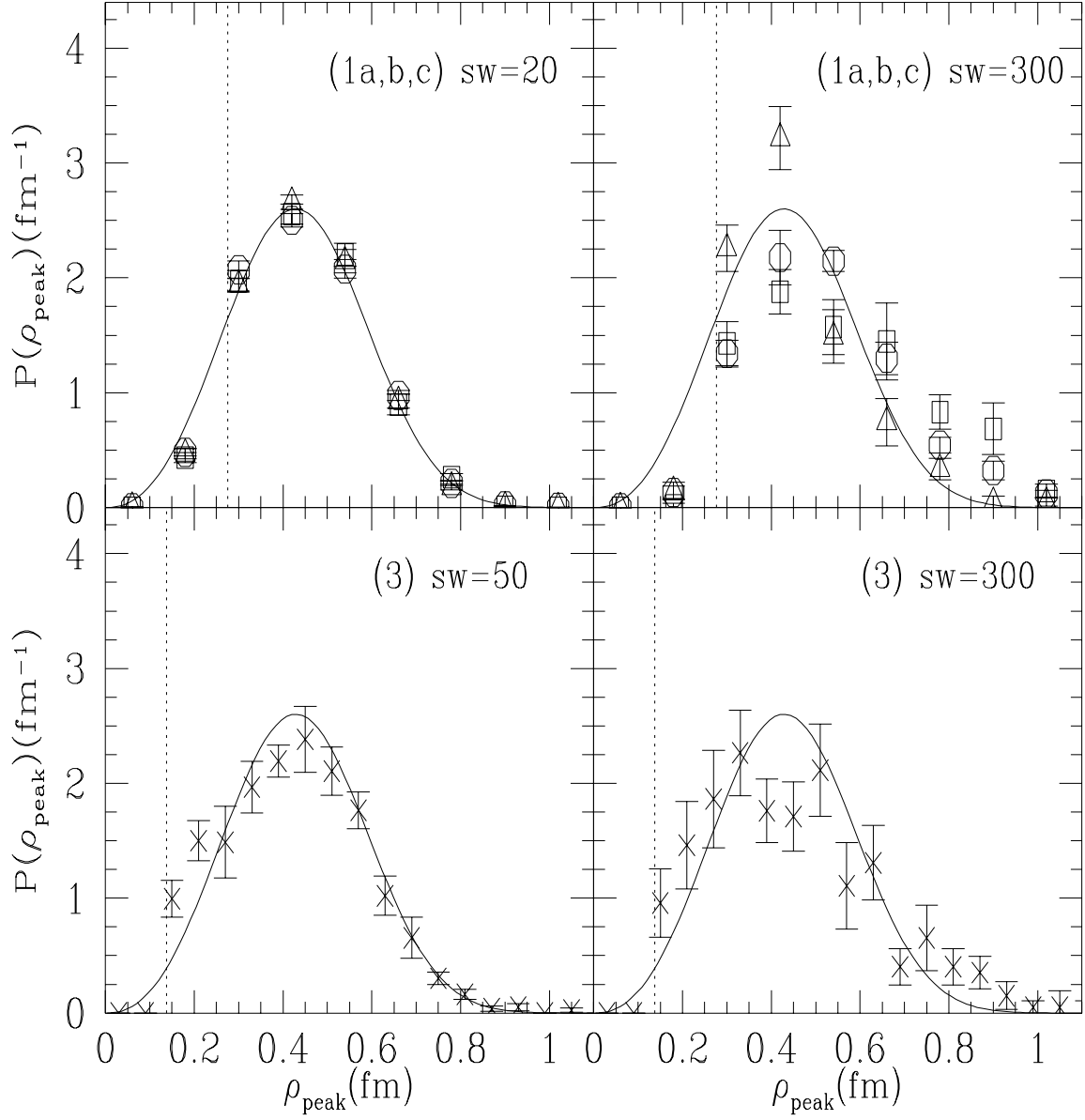}
\caption{Size distributions.
Crosses, squares, triangles, circles correspond respectively
to lattices (3), (1a), (1b), (1c) (see Table  \ref{t.dat}), the dotted vertical line is
the small size threshold $\tilde{\rho}_0=2.3 a$.
The curve in the figure corresponds to the fit eq. (\ref{e.rfit}).}
\label{f.siz}
\end{figure}
\newpage

\begin{figure}[htb]
\vspace{15cm}
\includegraphics{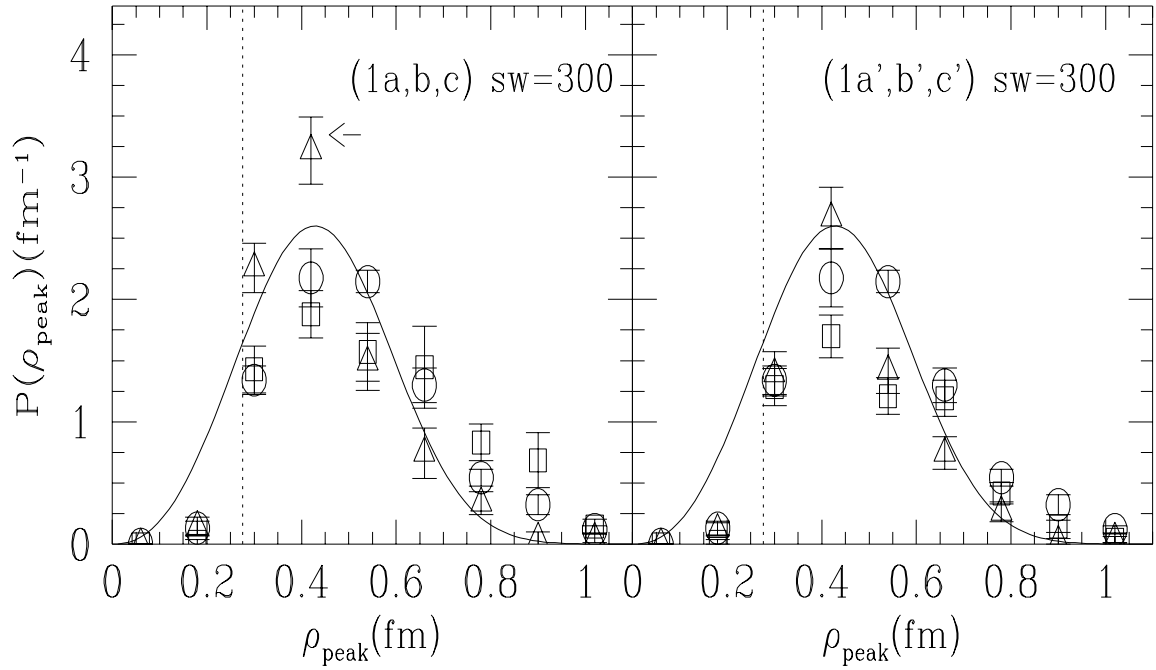}
\caption{Size distributions before and after subtracting configurations
with $\hat{S}=1$.
Squares, triangles and circles correspond to
(1a), (1b) and (1c) respectively (see Table  \ref{t.dat}), the dotted vertical line is
the small size threshold $\tilde{\rho}_0=2.3 a$.}
\label{f.fse}
\end{figure}

\vskip2cm
\begin{figure}[htb]
\vskip3cm
\vspace{14cm}
\includegraphics{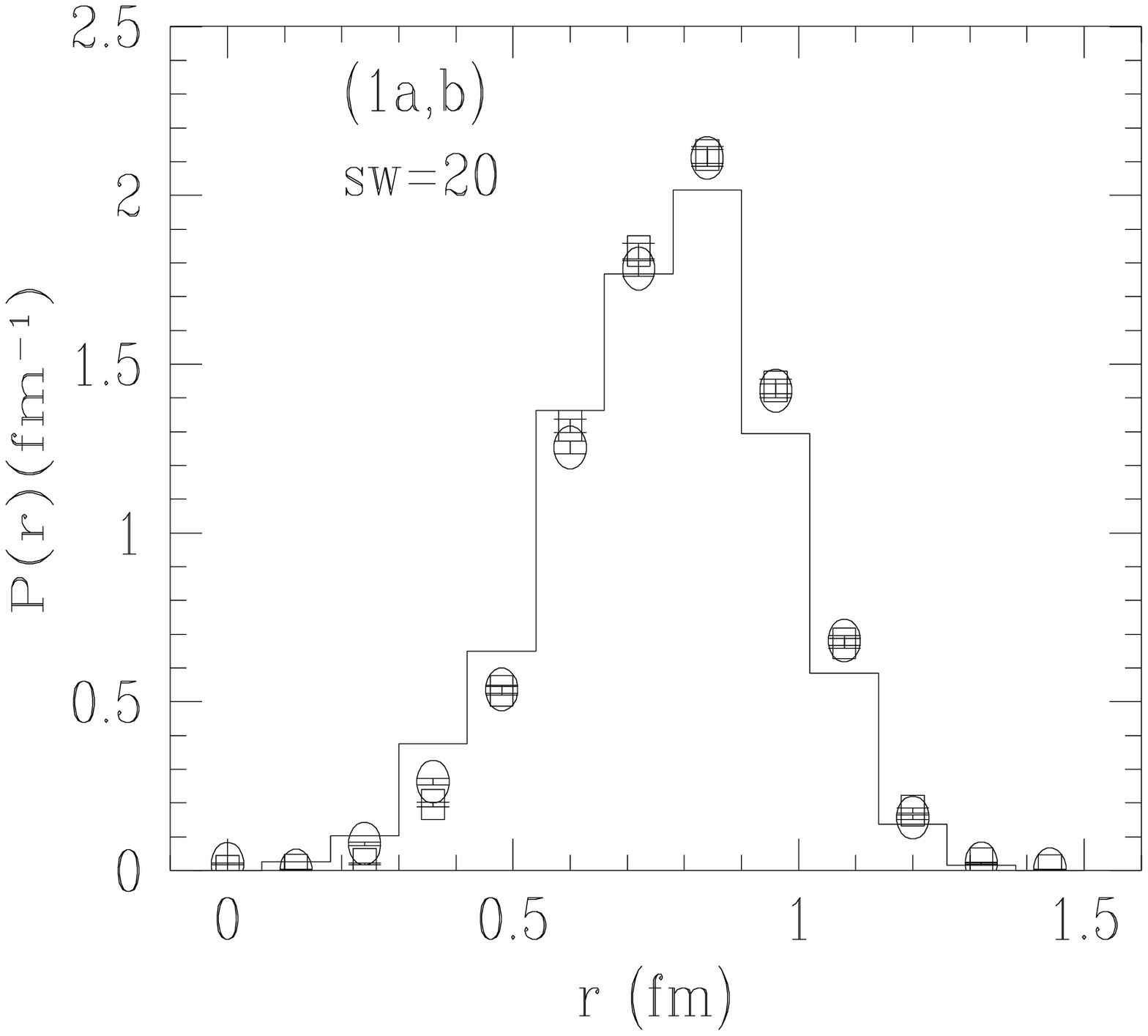}
\includegraphics{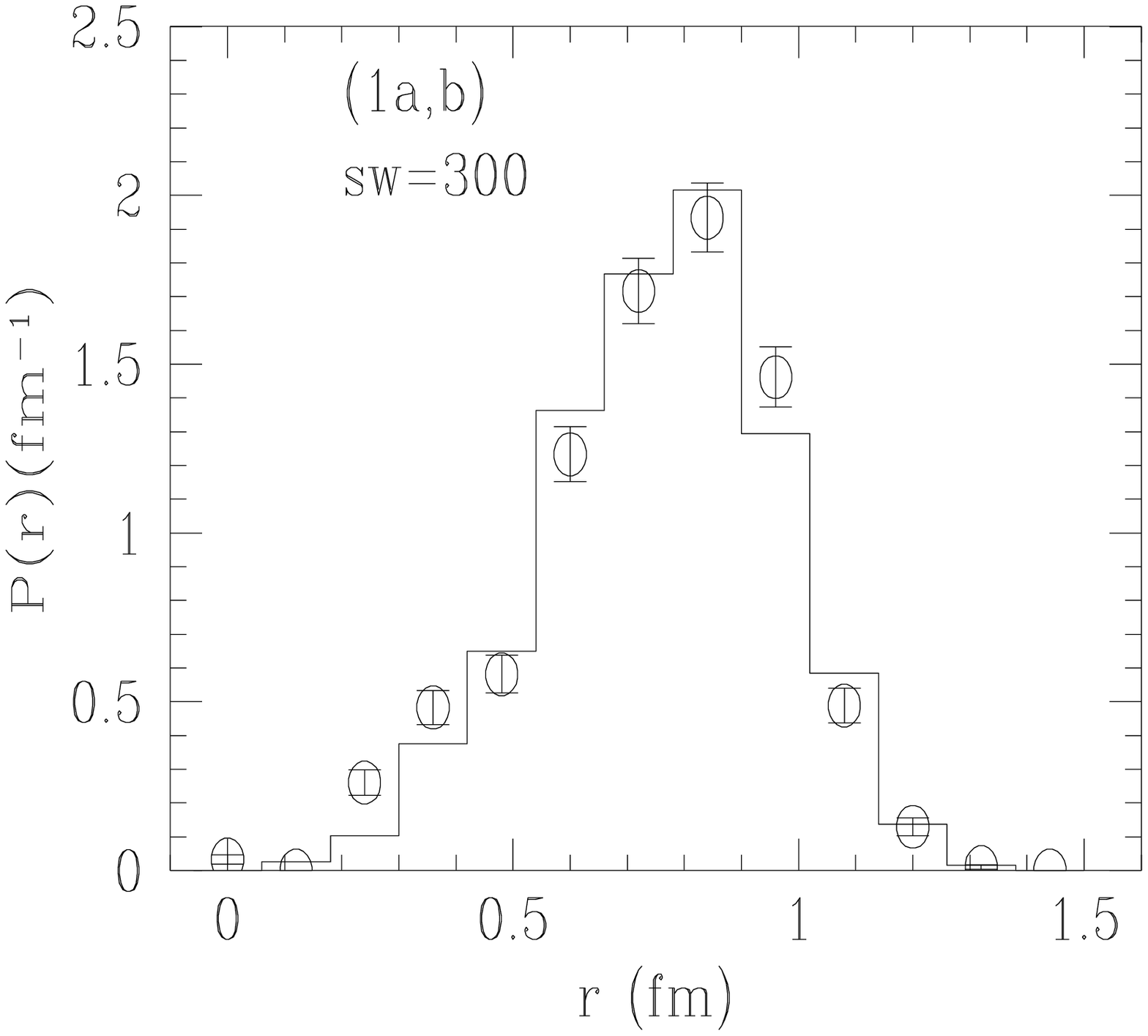}
\includegraphics{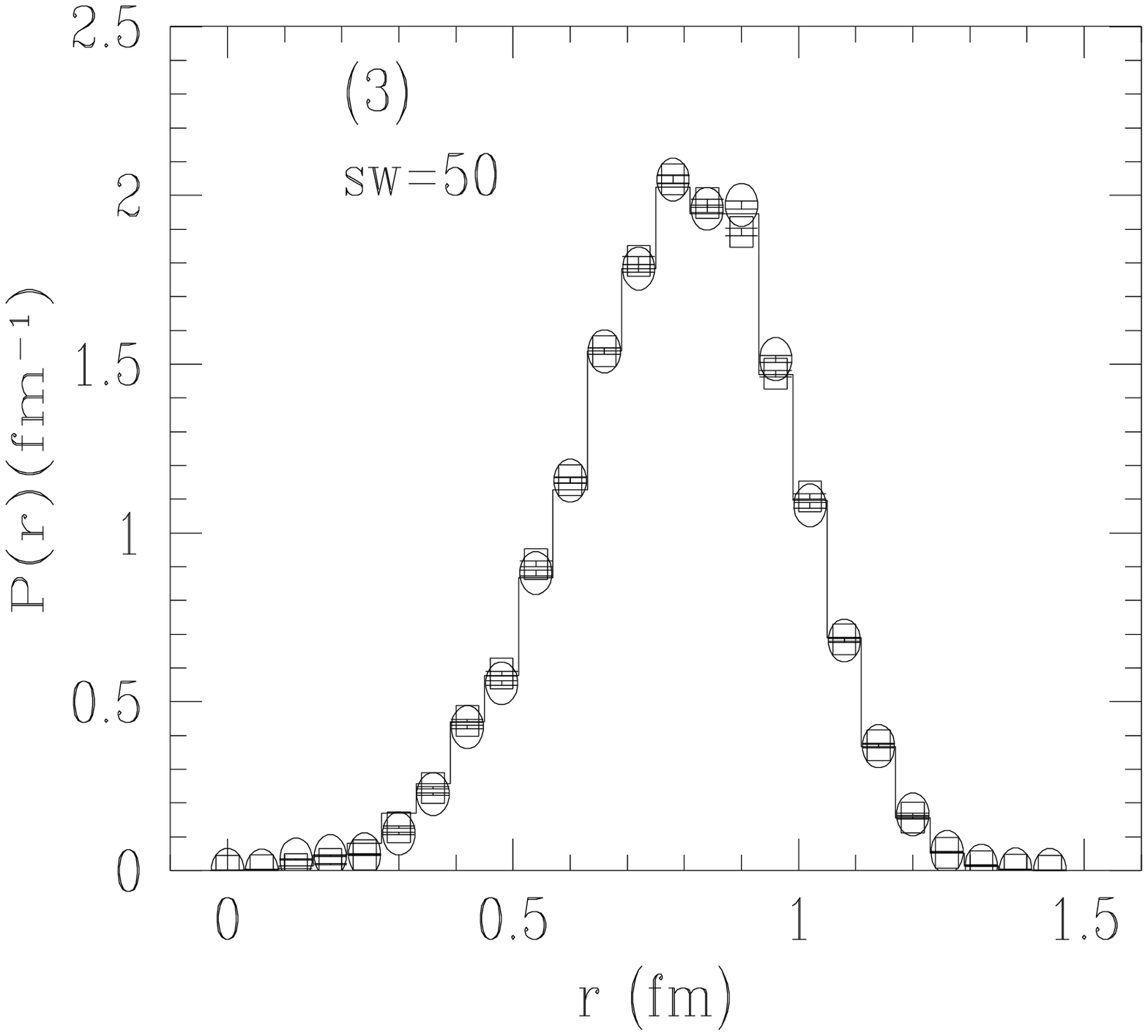}
\includegraphics{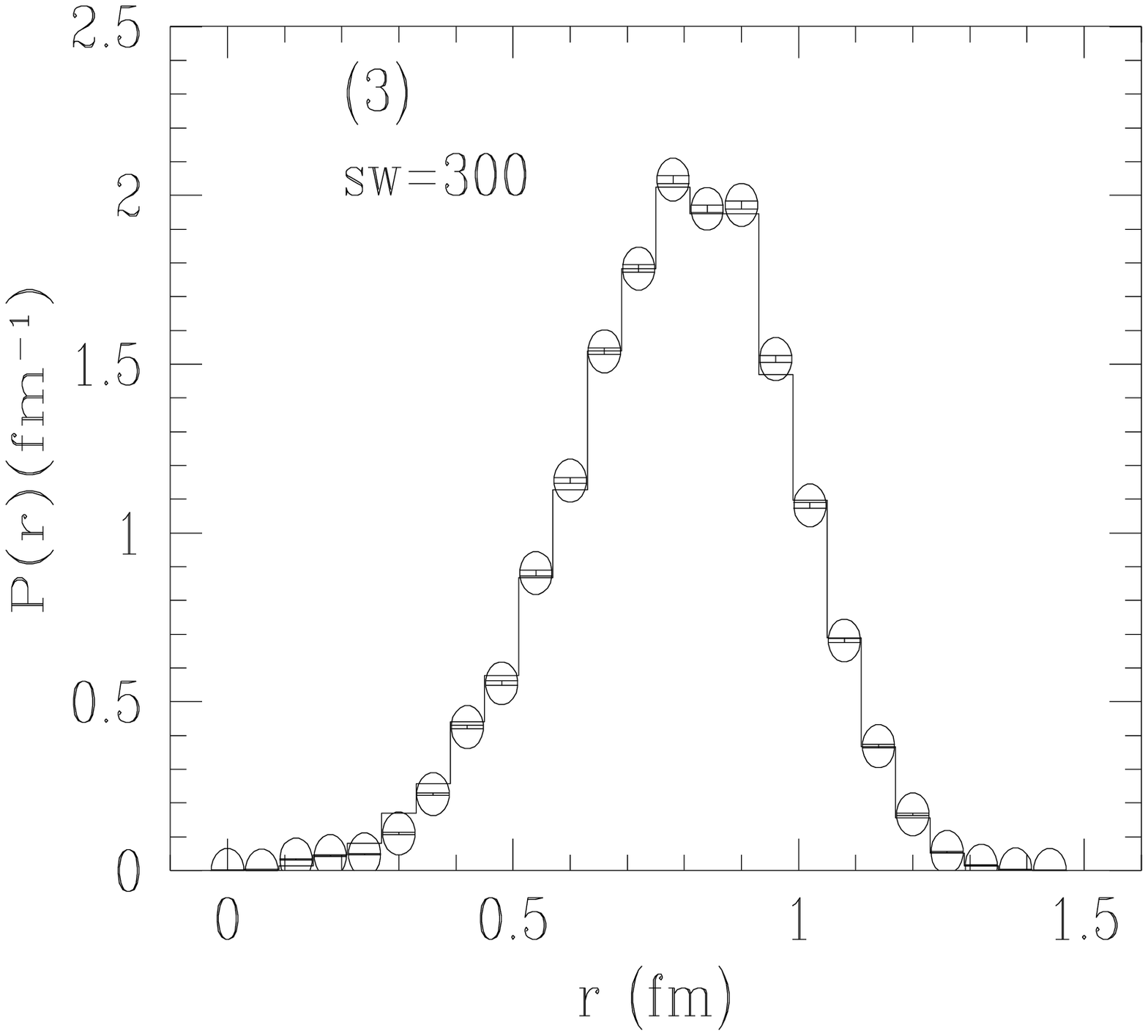}
\caption{Distribution of separation between instantons. Circles and 
squares represent separation between alike and 
dislike-charged peaks respectively. The solid-line histogram is the 4-dimensional volume factor
corresponding to a homogeneous distribution of (anti-)instantons over the lattice.
The upper plot is an average over (1a) and (1b)
lattices. The lower plot corresponds to the (3) lattice.}
\label{f.dis}
\end{figure}

\newpage
 
\begin{figure}[htb]
\vskip3cm
\vspace{14cm}
\includegraphics{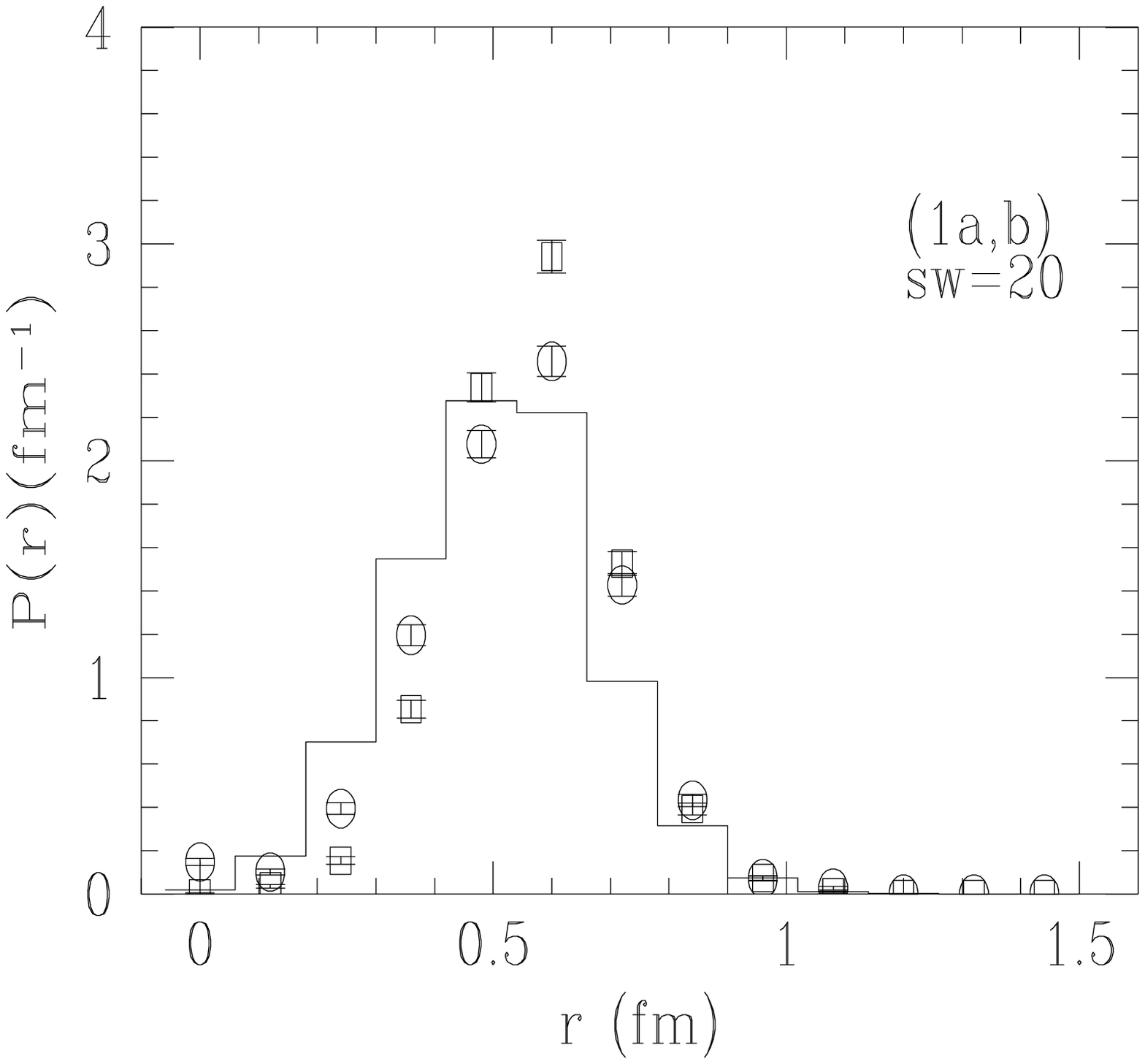}
\includegraphics{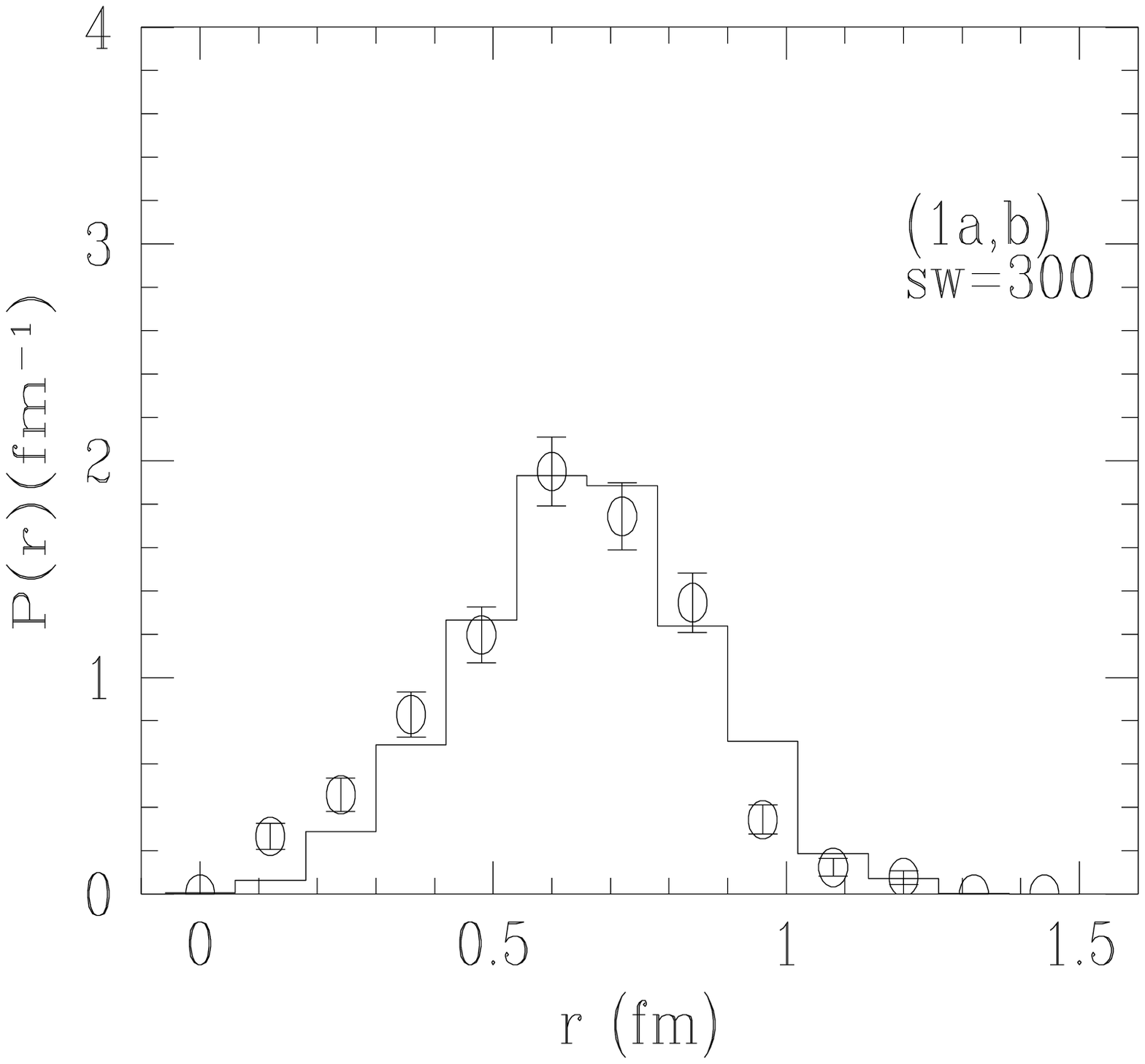}
\includegraphics{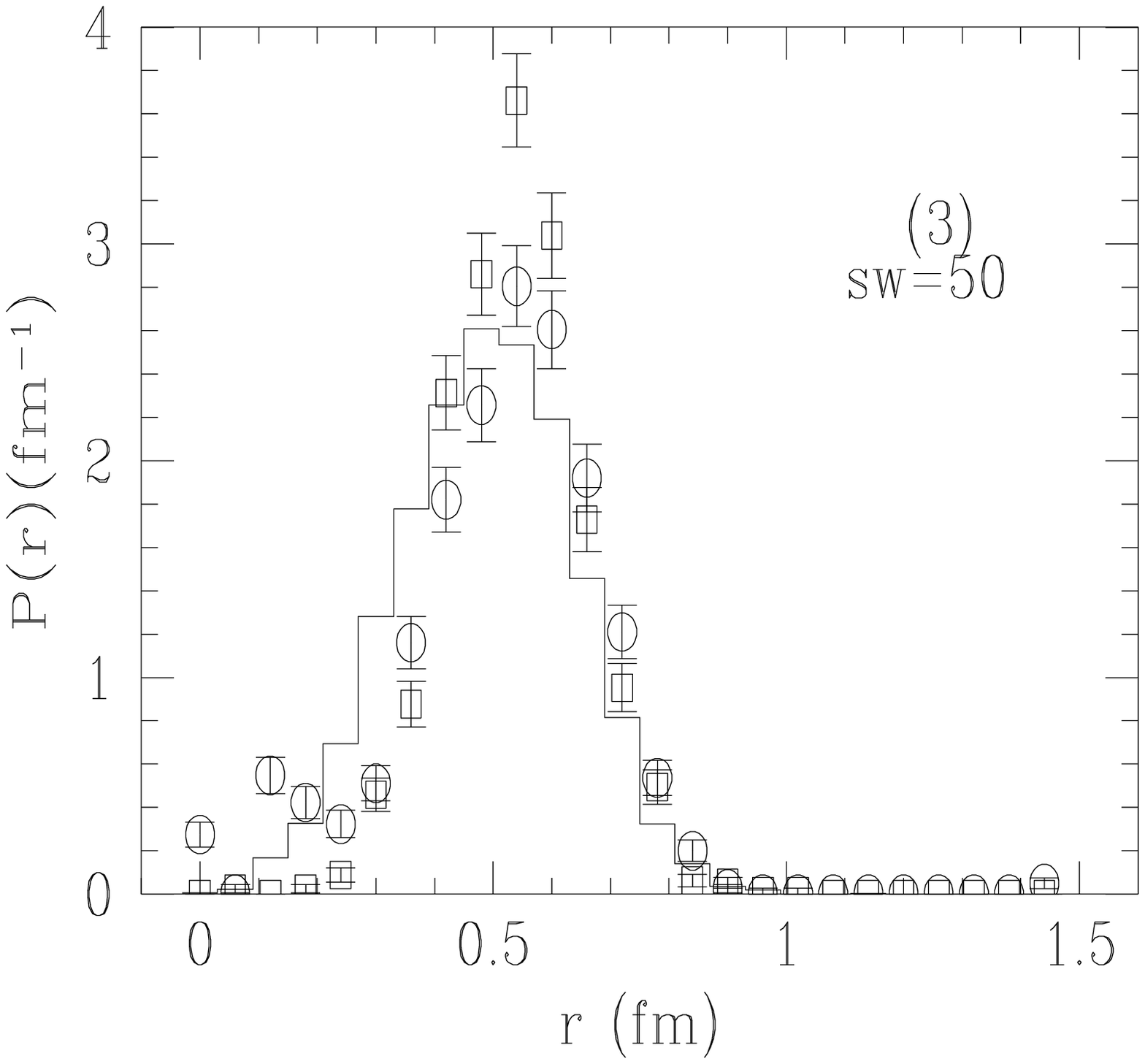}
\includegraphics{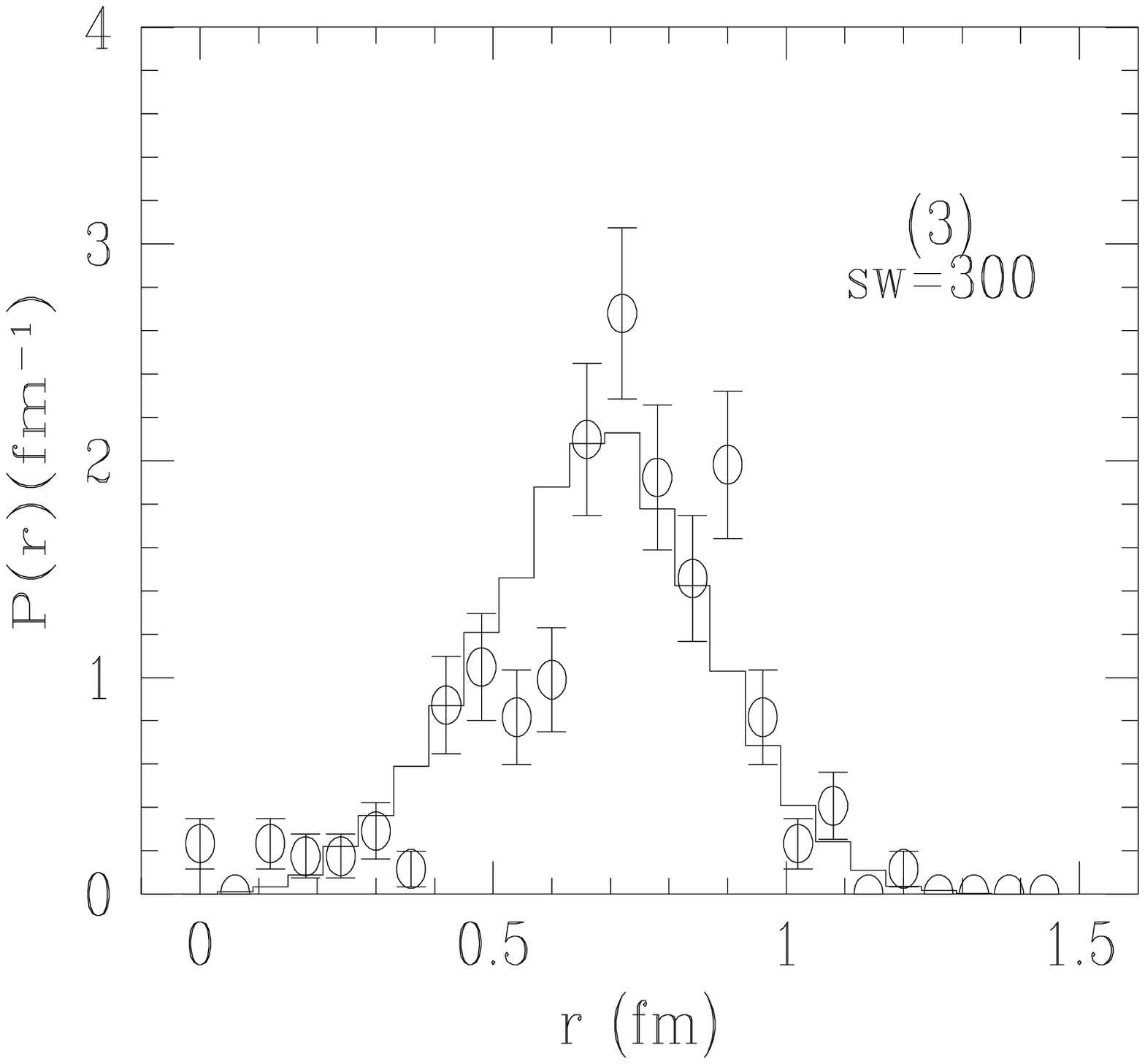}
\caption{Distribution of separation to the nearest  alike-charged (circles) or   
dislike-charged  (squares)  peak. The upper plot is an average over (1a) and (1b)
lattices. The lower plot corresponds to the (3) lattice.
The histograms in the figure represent simulations from homogeneous binomial
distributions as described in section~\ref{s.dis}.}
\label{f.dism}
\end{figure}

\newpage

\begin{figure}[htb]
\vspace{14cm}
\includegraphics{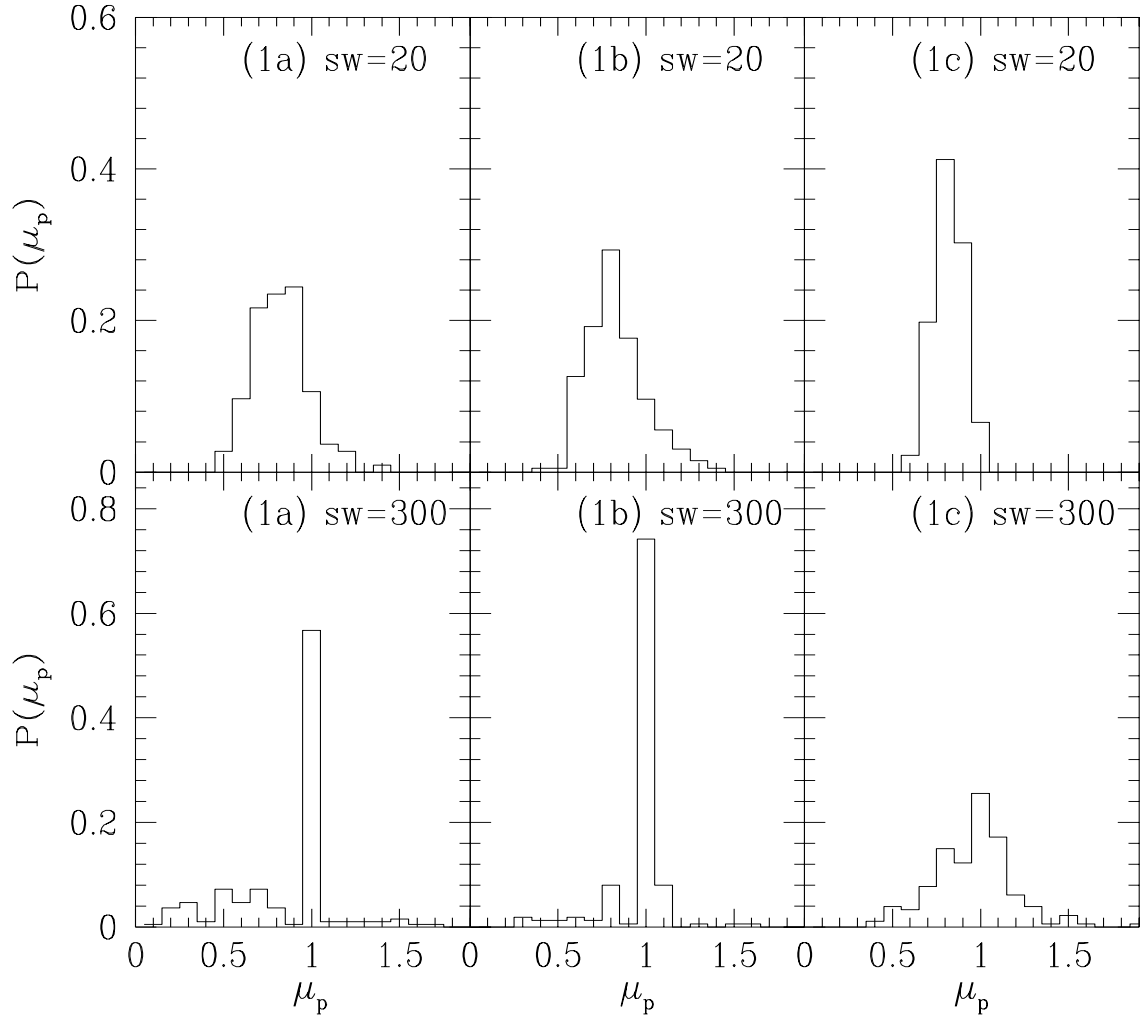}
\caption{Distribution of $\mu_{\rpp} = \hat{S}/N_{\rm peaks}$ with
$N_\rp$, the number of peaks in the energy density.}
\label{f.npeak}
\end{figure}

\newpage

\begin{figure}[htb]
\vspace{16cm}
\includegraphics{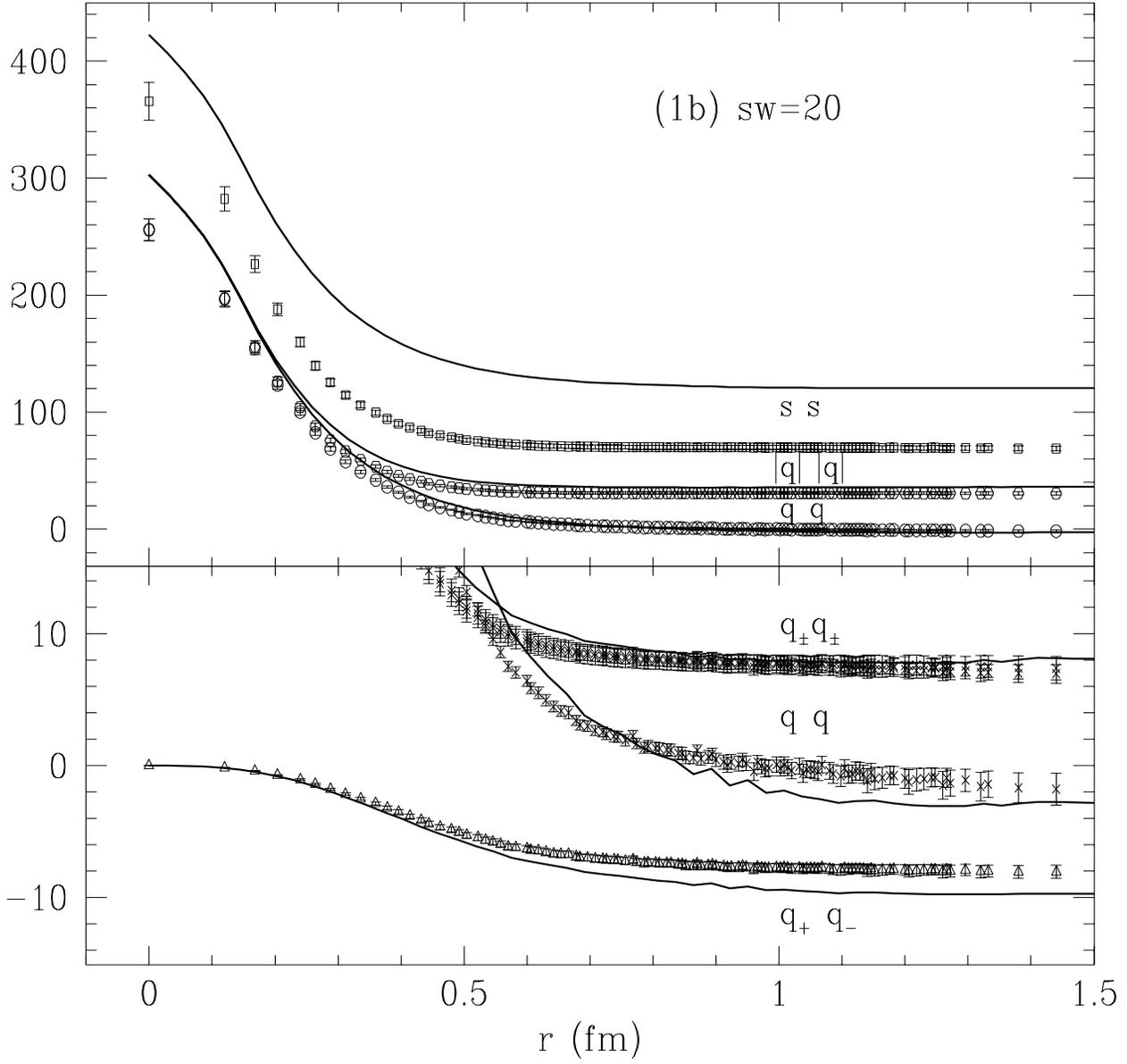}
\caption{Density-density correlations after 20 cooling sweeps
for lattice (1b). We plot 
 correlations of the dimensionless densities obtained from the continuum 
ones by multiplying with the physical volume of the lattice. In the upper
part of the plot we show $\langle \hat{s}(0)\hat{s}(r)\rangle$, $\langle q(0)q(r)\rangle$ and
$\langle |q(0)||q(r)|\rangle$ denoted as $s s$, $q q$ and $|q| |q|$ respectively.
In the lower part of the plot we have magnified the scale to show
a comparison between $\langle q(0)_\pm q_\pm(r)\rangle$ and $\langle q(0)_+ q_-(r)\rangle$
($q_\pm q_\pm$ and $q_+  q_-$ respectively in the figure).
The solid curves have been obtained from a homogeneous binomial
distribution  of instantons obeying the continuum ansatz as described in 
sections \ref{s.dis}, \ref{s.ia}.}
\label{f.qq2}
\end{figure}

\newpage

\begin{figure}[htb]
\vspace{16cm}
\includegraphics{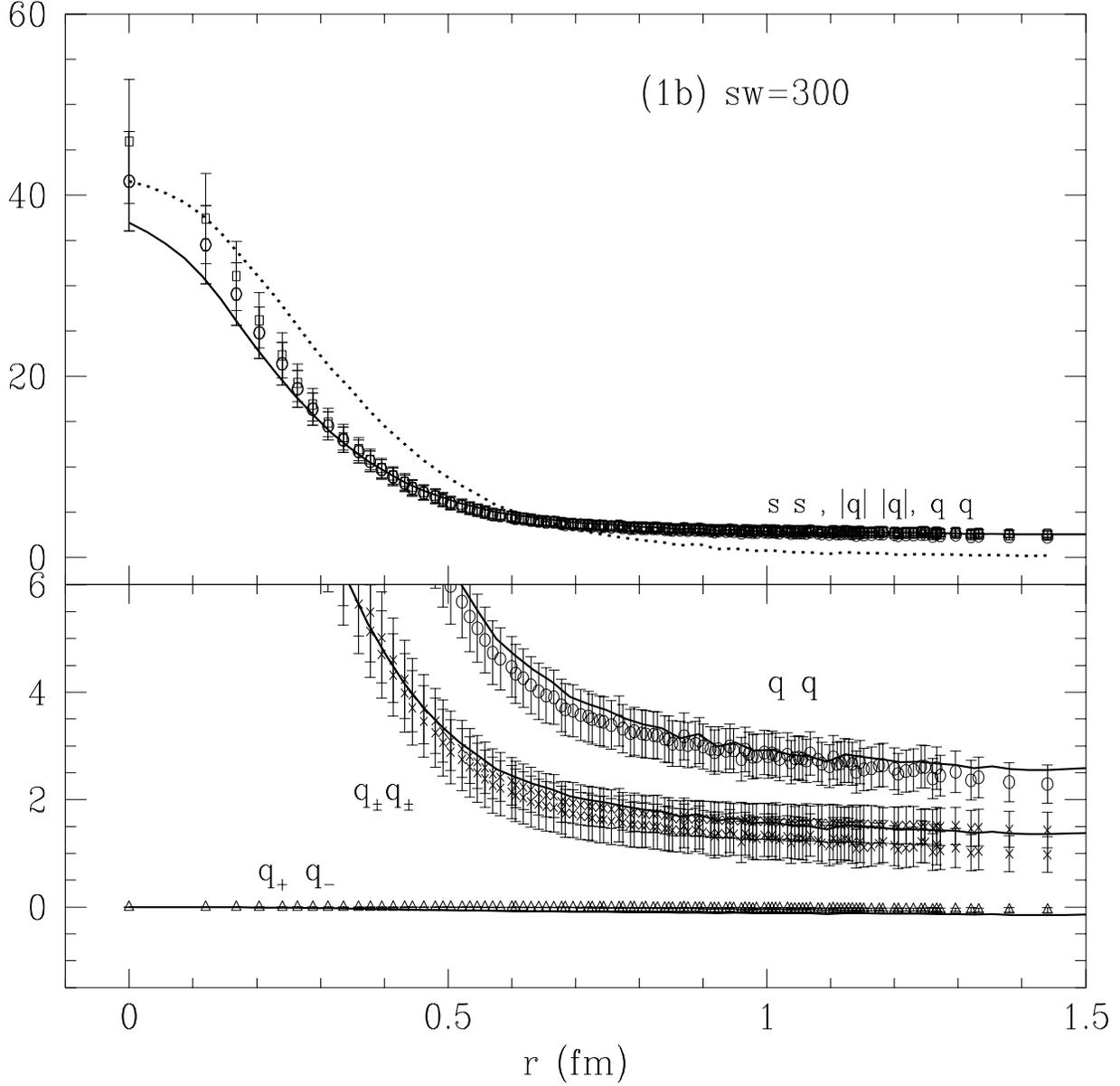}
\caption{Density-density correlations after 300 cooling sweeps
for lattice (1b). The dotted line 
represents the correlation obtained from  one
instanton of size $0.43$fm rescaled to match the measured value of the 
$\langle q(0)q(r)\rangle$ correlation at the origin.
The solid curves have been obtained from a homogeneous binomial
distribution  of instantons obeying the continuum ansatz as described in 
sections \ref{s.dis}, \ref{s.ia} (on the upper plot only the
curve for $q q$ appears since due to the small
number of I-A pairs the $s s$ and $|q| |q|$ curves practically coincide 
with this one).
Notation as in Fig. \ref{f.qq2}.}
\label{f.qq5}
\end{figure}

\begin{figure}[htb]
\vspace{16cm}
\includegraphics{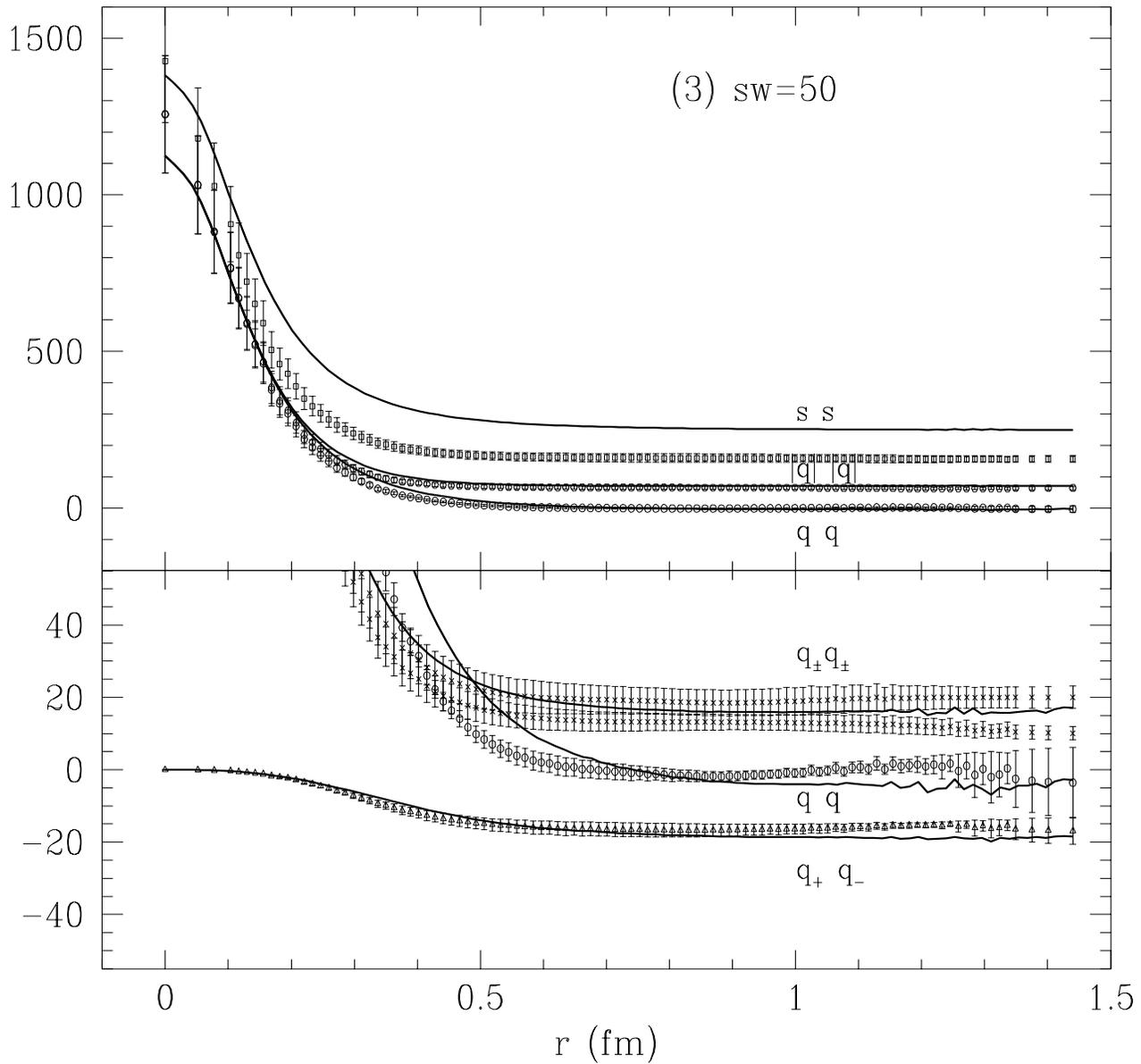}
\caption{Density-density correlations at 50 cooling sweeps for 20 configurations
from lattice (3).
The solid curves have been obtained from a homogeneous binomial
distribution  of instantons obeying the continuum ansatz as described in
sections \ref{s.dis}, \ref{s.ia}.
Notation as in Fig. \ref{f.qq2}.}
\label{f.qq24}
\end{figure}

\newpage
\begin{figure}[htb]
\vspace{16cm}
\includegraphics{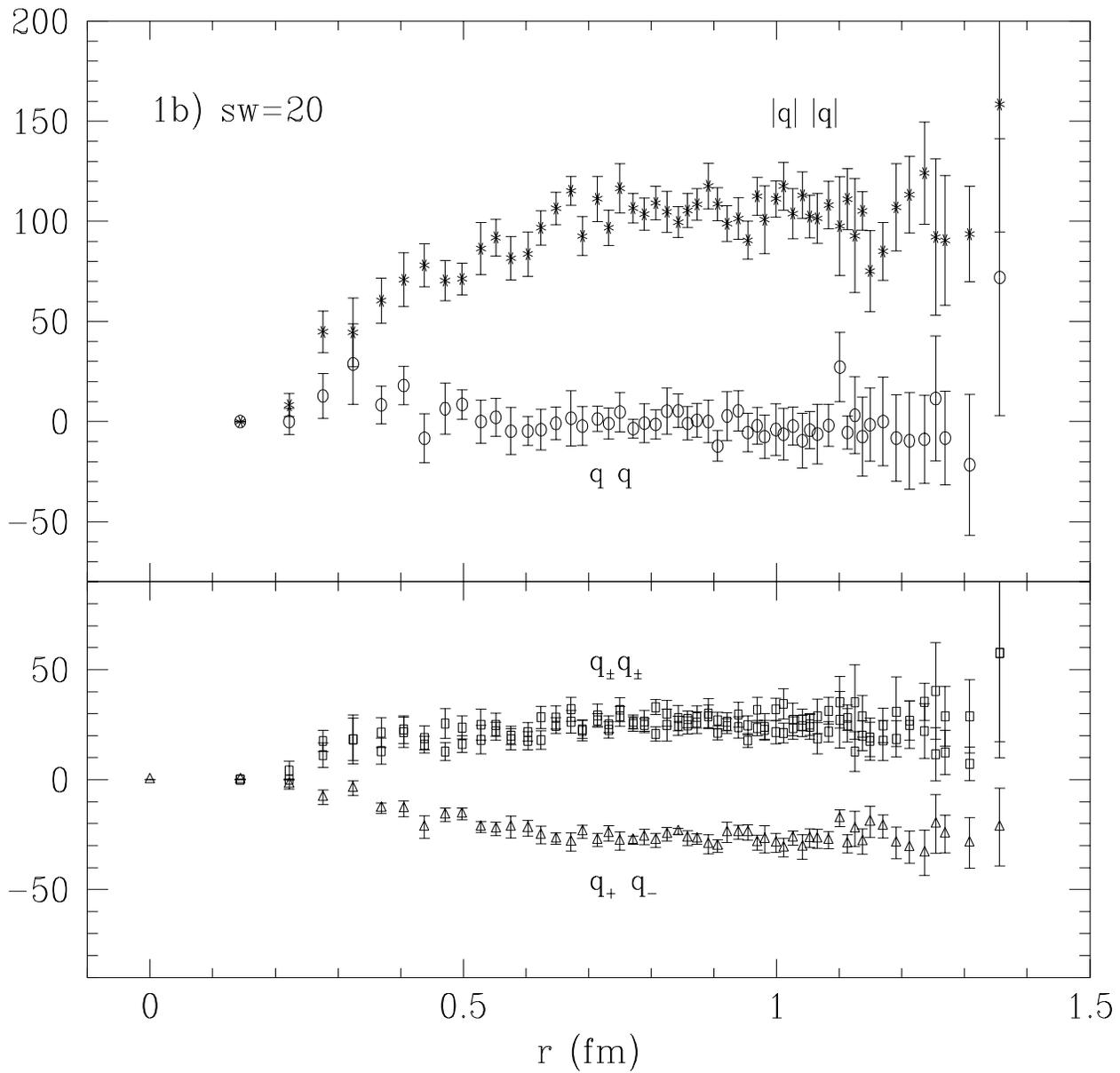}
\caption{Density-density correlations after 20 cooling sweeps 
for lattice (1b) obtained by placing a 
charge $Q = \pm 1$ at the location of the peaks of $q(x)$ and zero 
everywhere else.  Notation as in Fig. \ref{f.qq2}.}
\label{f.qqd}
\end{figure}

\end{document}